\documentclass[twocolumn,5p,monochrome]{elsarticle}
\biboptions{sort&compress}
\pdfoutput=1
\usepackage{graphicx}

\usepackage{amsmath}
\usepackage{algorithmic}
\usepackage{array}
\usepackage{float}
\usepackage{dblfloatfix}
\usepackage[caption=false,font=footnotesize]{subfig}
\usepackage{footnote}
\usepackage{url}
\usepackage{multirow}
\usepackage[usenames, dvipsnames]{color}
\usepackage{comment}
\usepackage{footnote}
\usepackage[figuresright]{rotating}
\usepackage{bbding}

\usepackage{hyperref}
\hypersetup{
    bookmarks=true,         
    unicode=false,          
    pdftoolbar=true,        
    pdfmenubar=true,        
    pdffitwindow=false,     
    pdftitle={Are Mobility Management Solutions Ready for 5G and Beyond?},    
    pdfauthor={Akshay Jain},     
    pdfsubject={},   
    pdfcreator={Akshay Jain},   
}
\usepackage{amsthm}
\makeatletter
\def\els@aparagraph[#1]#2{\elsparagraph[#1]{#2}}
\def\els@bparagraph#1{\elsparagraph*{#1}}
\makeatother
\makesavenoteenv{table}
\makesavenoteenv{tabular}

\begin{document}
\begin{frontmatter}

\title{Are Mobility Management Solutions Ready for 5G and Beyond?\tnoteref{mytitlenote}}
\tnotetext[mytitlenote]{This work has been supported in part by the EU Horizon 2020 research and innovation programme under grant agreement No. 675806 (5GAuRA), and by the ERDF and the Spanish Government through project RYC-2013-13029.}
\author[1]{Akshay~Jain \corref{mycorrespondingauthor}}
\ead{akshay.jain@upc.edu}

\author{Elena~Lopez-Aguilera\fnref{2}}

\author{Ilker Demirkol\fnref{f3}}

\cortext[mycorrespondingauthor]{Corresponding author}

\address{$^{1}$Department of Network Engineering, Universitat Polit\`{e}cnica de Catalunya BarcelonaTECH, Barcelona 08034, Spain.}

\begin{abstract}
Enabling users to move to different geograp\-hical locations within a network and still be able to maintain their connectivity and most essentially, continuity of service, is what makes any wireless network ubiquitous. Whilst challenging, modern day wireless networks, such as 3GPP-LTE, provision satisfactory mobility management (MM) performance. However, it is estimated that the number of mobile subscriptions will approximately touch 9 billion and the amount of data traffic will expand by 5 times in 2024 as compared to 2018.  Further, it is expected that this trend of exponential growth will be maintained well into the future. To cope with such an exponential increase in cellular traffic and users alongside a burgeoning demand for higher Quality of Service (QoS), the future networks are expected to be highly dense and heterogeneous. This will severely challenge the existing MM solutions and ultimately render them ineffective as they will not be able to provide the required reliability, flexibility, and scalability. Consequently, to serve the 5G and beyond 5G networks, a new perspective to MM is required. Hence, in this article we present a novel discussion of the functional requirements from MM strategies for these networks. We then provide a detailed discussion on whether the existing mechanisms conceived by standardization bodies such as IEEE, IETF, 3GPP (including the newly defined 5G standards) and ITU, and other academic and industrial research efforts meet these requirements. \textcolor{red}{We accomplish this via a novel qualitative assessment, wherein we evaluate each of the discussed mechanisms on their ability to satisfy the reliability, flexibility and scalability criteria for future MM strategies. We then present a study detailing the research challenges that exist in the design and implementation of MM strategies for 5G and beyond networks. Further, we chart out the potential MM solutions and the associated capabilities they offer to tackle the persistent challenges. We conclude this paper with a vision for the 5G and beyond MM mechanisms.}     
\end{abstract}

\begin{keyword}
5G, Beyond 5G, 6G, Mobility Management, SDN, Meta-Surfaces.
\end{keyword}

\end{frontmatter}


\section{Introduction}

Future wireless networks define a very challenging environment for mobility management (MM) solutions, due to the significant increase in density (in terms of both users and deployed access points), in heterogeneity (given the various radio access technologies (RATs) supported), as well as in programmability (the network as well as the environment can be programmable). To achieve an ubiquitous network service in such challenging environments, it is critical to devise effective MM strategies that facilitate seamless mobility by allowing users to traverse through the network without losing connectivity and service continuity. 

One of the traditional approaches for allowing applications to serve a user in mobile scenarios has been to maintain network connectivity through handovers \textcolor{red}{based on criteria such as Radio Signal Strength Indicator (RSSI), Signal to Interference and Noise Ratio (SINR), Reference Signal Received Quality (RSRQ), Reference Signal Received Power (RSRP), etc}. However, in addition to the signal quality parameter centric handovers, modern day applications necessitate that other parameters such as available core network bandwidth, End-to-End (E2E) latency, backhaul bandwidth and backhaul reliability \cite{Sutton2018} are also taken into consideration. Moreover, maintaining Quality of Service (QoS), e.g., provisioning service continuity, link continuity, required bit-rate and latency, during mobility scenarios has been one of the primary objectives for novel MM mechanisms. Multiple strategies to satisfy such QoS criteria such as service migration \cite{Machen2018}, service replication \cite{Frangoudis2018}, path reconfiguration \cite{Yang2016}, etc., have been proposed by the research community. MM solutions for 5G and beyond networks are also expected to ensure E2E connectivity and session continuity through the maintenance/preservation of IP address of the user towards the core network entity that provisions the service for the corresponding user.

\begin{figure*}
	\centering
	\includegraphics[scale = 0.45]{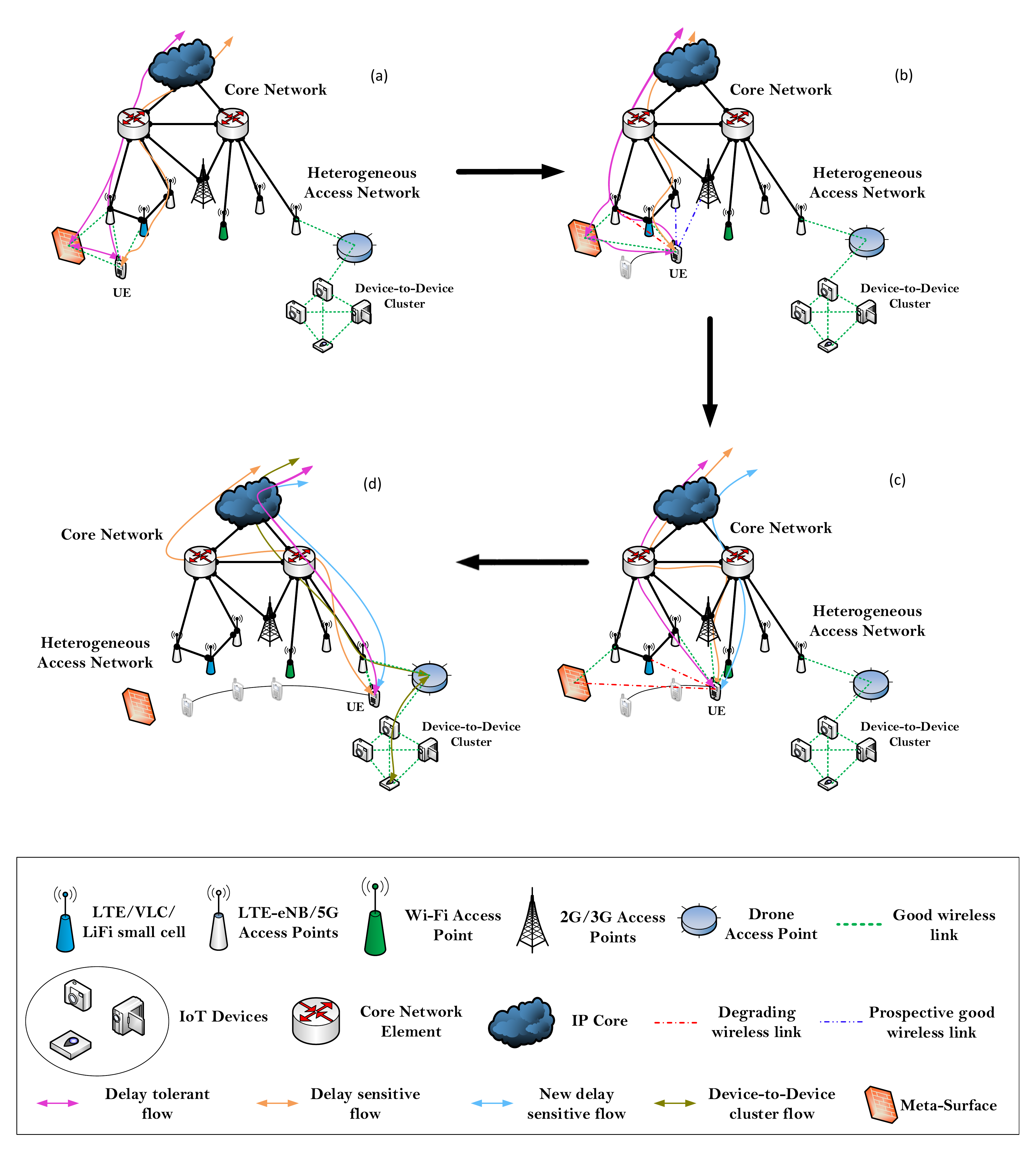}
	\caption{An illustrative 5G and beyond network mobility scenario.}
\end{figure*}

\textcolor{red}{To motivate further, we consider} an illustrative example of the future mobility scenario as presented in Figure 1. It shows the extraordinary nature of complexity that the future networks will present for MM. As shown in Figure 1(a), a mobile user equipment (UE) is connected to multiple RATs (5G Access Point (AP)/ Long Term Evolution (LTE) eNode B (eNB)/visible light communications (VLC) and Light Fidelity (LiFi) small cells \cite{Boulogeorgos2018, Chowdhury2018, Zhang2019}, etc.), while having a delay tolerant and a delay sensitive application datastream (flows) with distinct QoS profiles. Also, the AP through which the delay tolerant flow is being served to the user has a good wireless link with a meta-surface in the vicinity. While, traditionally the environment between the user and an AP is considered as an adversary in all the generations of mobile communications, including 5G, in beyond 5G (B5G) networks the environment will be programmable and hence, will be an ally by provisioning favorable transmission channels \cite{Basar2019, Renzo2019, Chung2011}. These favorable channels will essentially consist of reflected signals, the phases and polarizations of which will be adjusted by thin (but electrically significant) surfaces, also known as meta-surfaces, so that they interfere constructively at the receiver \cite{Renzo2019, Basar2019}. In addition to the meta-surfaces, future networks will also consist of mobile APs such as drones, as shown in Figure 1(a). Note that, the density of meta-surfaces and drone APs will also be extremely high in future networks. Further, in the scenario illustrated, we consider the use case wherein the drone AP is servicing a device-to-device (D2D) cluster, and connecting it to the core network through one of the ground based APs. The D2D cluster over the course of its existence does not generate packets as frequently as the other users, since the cluster devices mainly host Internet of Things (IoT) applications.


Next, in Figure 1(b), as the user moves, it starts to register wireless links with better signal quality from other APs as compared to those it is already associated to. It is imperative to state here that, the APs can be from the same or different network operators. Henceforth, a careful and efficient RAT and AP selection for each flow will be necessary as part of the future MM mechanisms. It is interesting to observe that while the AP used for serving the delay tolerant flow in Figure 1(a) no longer has a good link quality, through the meta-surfaces and their programmable nature it still has a good wireless link to the user and hence is able to serve it.

Following the new RAT/AP association, flows pertaining to the user are redirected through the most optimal path. Novel MM mechanisms that aim to service the 5G and B5G networks will require efficient route optimization methods to perform the same. Additionally, the MM mechanisms will also need to implement IP forwarding so as to ensure E2E link continuity. In Figure 1(c) we then observe that as the user moves further, the RAT/AP selection and optimal routing methods are continually implemented. Further, when a new application request is generated, as seen in Figure 1(c), an appropriate RAT and AP for the given flow is selected alongside the route that satisfies the requested QoS. Lastly, in Figure 1(d), it can be seen that alongside the user's flows, the D2D cluster's flows are also being serviced by network. However, the D2D cluster is firstly serviced by a drone AP, which then relays information to/from the ground based APs. These ground based APs assist in serving the data flows generated from the devices in the D2D cluster by relaying the data to the relevant servers in the core network.

Given the complexity of the scenario presented in Figure 1, it is evident that no single MM mechanism will form the solution to all the possible situations and scenarios that will be prevalent. And, although current MM mechanisms propose methods for careful RAT and AP selection, IP packet forwarding, route optimization, and session management, a more than 10-fold increase in user density coupled with the heterogeneity in flow types and network will extremely limit their capabilities, as explained in the subsequent sections in detail. New user applications such as Augmented Reality, Virtual Reality, Vehicle-to-Everything (V2X), etc., will present very restrictive delay requirements, exceptionally high reliability and bandwidth requirements \cite{Parvez2018}, that will consequently severely challenge the capabilities of current MM strategies. Further, the radio access network (RAN) technologies themselves are expected to undergo important transformation in the future networks given the significant interest in VLC, LiFi, etc., \cite{Boulogeorgos2018, Chowdhury2018}. Whilst both LiFi and VLC, being TeraHertz (THz) bandwidth technologies, enable near Terabits per second (Tbps) speeds, they are significantly impaired by the environment. This consequently has significantly more detrimental effects on the user QoS during mobility scenarios, which we will discuss in further detail in the later sections.

Also, owing to the telecom operators' desire to serve more industry verticals, a new set of mobility patterns will emerge. For example, a platoon of vehicles moving coherently together, vehicles disbanding from one platoon to join another, ultra-fast moving users (in excess of 500 km/h), moving access points (such as those on drones \cite{Sekander2018}), etc., thus introducing another dimension to the MM problem. Henceforth, the ability to serve devices with mobility patterns that will be more diverse and challenging as compared to current day network scenarios, will be a significant challenge towards the design, development and deployment of 5G and beyond MM mechanisms. An additional yet significant challenge will be to manage and potentially reduce the control plane (CP) signaling load \cite{Jain2019} due to mobility events.

Thus, a fresh perspective, wherein MM solutions are decentrali\-zed and flexible, can support multiple use cases simultaneously and account for the various other radical changes in 5G and B5G networks \textcolor{red}{with reliability}, is required. Note that, decentralization will permit MM mechanisms to service the exponentially increasing number of users coupled with different mobility profiles (e.g., static IoT devices and users in high-speed trains). On the other hand, flexibility will allow them to adapt to the user, network and/or environment context (e.g., QoS, user mobility profile, network load, flow types, meta-surfaces, etc.). \textcolor{red}{Additionally, reliability will aid in provisioning seamless mobility as well as in satisfying the ultra-reliable criterion for future wireless network applications.}

References \cite{Akyildiz2015} and \cite{Andreev2014} aim to provide new MM strategies via Software Defined Networking (SDN) based MM and multi-RAT mobility. However, they do not elaborate on the myriad challenges that future MM mechanisms will encounter, such as time complexity, signaling overhead, etc. \textcolor{red}{Similarly, while in \cite{Fan2016} MM strategies, such as advanced cell association, group handovers, etc., have been discussed to address the heterogeneity in the mobility patterns and profiles that will arise in 5G, they fall short in addressing the challenges such as core network signaling, complexity, etc., that 5G and beyond MM solutions will face.} Further, surveys such as \cite{Ferretti2016} and \cite{Zekri2012} are restricted to the current network architecture, and hence, fail to provide a MM perspective for 5G and beyond networks. In addition, while \cite{Zhang2019} aims to provide insights into the requirements, architecture and key technologies for B5G networks, it does not address the critical issue of MM in B5G networks. Hence, to the best of our knowledge, no study has ever provided a comprehensive view of the functional requirements, challenges and potential solutions with regards to the future MM strategies, essential to realizing the future networks. \textcolor{red}{We now list the contributions of this paper, which aim to address these aforementioned gaps, as follows:}
\begin{itemize}
    \item[\textcolor{red}{1.}] \textcolor{red}{We present a novel discussion on the functional requirements and design criteria for 5G and Beyond MM mechanisms.} 
    \item[\textcolor{red}{2.}] \textcolor{red}{We develop a novel qualitative analysis for the legacy mechanisms as well as the current state of the art MM mechanisms on the basis of reliability, flexibility and scalability, towards their utility for 5G and beyond wireless networks.}
    \item[\textcolor{red}{3.}] \textcolor{red}{We provide a novel classification of the current state-of-the-art mechanisms based on where they are implemented or create an impact within the network, i.e., core network (CN), access network (AN) and extreme edge network. Additionally we also provision a mapping of these classifications onto the 5G service based architecture (SBA) defined by 3GPP \cite{3GPP2020}, which will consequently assist to indicate explicitly the gaps that exist currently.} 
    \item[\textcolor{red}{4.}] \textcolor{red}{We then provide the first discussion in literature with regards to how the current state-of-the-art strategies will fare towards MM for potential B5G solutions envisioned.}
    \item[\textcolor{red}{5.}] \textcolor{red}{Following the discussions and qualitative analysis we have elucidated the various challenges that the design and development of future MM mechanisms will face.}
    \item[\textcolor{red}{6.}] \textcolor{red}{We then provide a discussion on the potential strategies that will help them overcome these persistent challenges. We accompany these discussions with a novel mapping between the potential strategies and the aforementioned challenges that they will help resolve.}
    \item[\textcolor{red}{7.}] \textcolor{red}{Lastly, we develop and provision a novel and unified vision for the 5G and beyond MM solution.}
\end{itemize}

\noindent \textcolor{red}{The rest of this paper is organized as follows: Section 2 presents the functional requirements and design criteria for the 5G and beyond MM mechanisms. Section 3 defines the criteria for the qualitative analysis as well as the parameters that govern the fulfillment of these criteria.  Section 4 presents the novel qualitative analysis for the legacy mechanisms and establishes their pros and cons for 5G and beyond MM. Section 5 introduces a similar analysis for the current state of the art mechanisms as well as their utility towards the MM solutions fr future networks. Section 6 then presents the persistent challenges, the potential strategies that will assist in resolving these challenges whilst aiming to satisfy the requirements defined in Section 2, and the proposed framework for 5G and beyond MM. We then conclude this paper in Section 7.}

\section{5G and Beyond MM: Functional Requirements and Design Criteria}

Future wireless networks, in addition to being dense, heterogeneous and extensively programmable, will serve multiple industry verticals as well as accommodate multiple tenants on the same network infrastructure \cite{Renzo2019,Rost2016}. These transformations, some of which are being discussed by the research community \cite{Basar2019, Akyildiz2016}, represent a paradigm shift from the current network architecture design. As a consequence, MM mechanisms need to be re-evaluated and/or re-designed. For this, we first present the functional requirements of MM mechanisms for future wireless networks in Table 1, based on the characteristics we derive from the current and future network scenarios. 


From Table 1, it can be observed that the MM solutions for 5G and beyond networks will have to adapt and evolve, so as to be able to serve the future wireless networks efficiently. As seen from the table, MM solutions will need to be redesigned so that they are flexible, scalable and reliable to ensure the requested QoS and seamless mobility. Apart from these requirements, there are certain criteria that will impact the design and development of future MM solutions. \textcolor{red}{Consequently, in the following text we present an insight into these myriad design criteria and their impact on 5G and beyond MM.} 

\begin{table*}
    \caption{Functional Requirements from 5G and beyond MM} 
    \centering
    \begin{tabular}{|p{0.55cm}|p{4cm}|p{7cm}|p{5.5cm}|} \hline
        \textbf{Req \#} &\textbf{Current Scenario} & \textbf{5G and Beyond Scenario} & \textbf{Resulting MM Functional Requirement} \\ \hline
        \textcolor{red}{\textbf{R1}} & Single RAT connectivity & UE connected to multiple RATs & Provision support for multi-RAT MM as well as efficient RAT selection methods.\\ \hline
         \textcolor{red}{\textbf{R2}} & UEs with predominantly mobile broadband applications request MM support & UEs with enhanced broadband (eMBB), massive machine type communications (mMTC) and Ultra-reliable low latency communication (URLLC) applications will request MM support. These applications will have different QoS requirements \cite{Elayoubi2016}. For example: minimum data rate, latency, reliability, etc. & Provide MM support based on context, i.e., based on application requirements, user mobility, network conditions, etc. \\ \hline
         \textcolor{red}{\textbf{R3}} & Density of UEs in the current scenario is $10^5 devices/km^2$ \cite{ITU2015} & Density  of UEs in 5G and beyond will be $\geq 10^6 devices/km^2$ \cite{ITU2015} & MM mechansims should be able to scale and provision support for the increasing user density \\ \hline
         \textcolor{red}{\textbf{R4}} & Network is vendor driven \cite{Habibi2019} & Network is softwarized \cite{Habibi2019} & MM solutions should evolve to utilize the benefits provided by softwarized 5G and beyond enablers such as SDN, Network Function Virtualization (NFV), etc. \\ \hline
         \textcolor{red}{\textbf{R5}} & Network is predominantly ground based with static radio towers & APs and relay stations may be carried on drones in 5G and beyond networks \cite{Khawaja2019, Li2019} & MM solutions for 5G and beyond networks should support mobility of both UEs and APs \\ \hline
         \textcolor{red}{\textbf{R6}} & 4G, 3G and 2G are standardized and the MM protocols provision support for all these devices & 5G and beyond networks and devices will be gradually rolled out. They are fundamentally different from 4G, 3G and 2G networks & Backwards compatibility to support legacy devices will be needed from MM solutions for 5G and beyond. \\ \hline
         \textcolor{red}{\textbf{R7}} & Sub 6 GHz is the frequency range for data transfer & Sub 6 GHz, millimeter Wave (mmWave) \cite{Akyildiz2016}, Terahertz communication \cite{Boulogeorgos2018, Chowdhury2018} will be utilized in 5G and beyond networks & Increased robustness, given that VLC and mmWave will be significantly impacted by the environment, thus challenging seamless mobility in 5G and beyond networks. \\ \hline
         \textcolor{red}{\textbf{R8}} & Finest granularity of tracking and localization is $<50 m$ \cite{Wymeersch2017} & Finest granularity of tracking and localization is a beam ($<10 cm$) \cite{Wymeersch2017} & MM solutions should evolve to utilize the advanced level of granularity to provision better mobility and tracking performance in dense urban or high speed scenarios \\ \hline
         \textcolor{red}{\textbf{R9}} & The complexity is driven mainly via user requirements in a homogeneous network & The complexity in 5G and beyond networks is a combination of different user types, different QoS requirements, heterogeneous RAT scenarios, heterogeneous backhaul scenarios \cite{Jaber2016b} and ultra dense nature of the network & MM mechanisms need to ensure adequate flexibility (they should accommodate for the increased heterogeneity) and tractable solutions (fast and low computational complexity) with well managed power consumption for the increased network complexity \\ \hline
         \textcolor{red}{\textbf{R10}} & Requested services and data is always hosted in the IP Multimedia Subsystem (IMS) core & Requested services and data in 5G and beyond networks can now be hosted at the network edge, through MECs \cite{Habibi2019} & MM mechanisms should provide adequate Service Migration \cite{Machen2018}/ Service Replication \cite{Frangoudis2018} support to ensure the required QoS from the applications \\ \hline
         \textcolor{red}{\textbf{R11}} & Support for mobility up to 350 km/h & Support for mobility up to and beyond 500 km/h proposed \cite{ITU2015} & MM solutions need to ensure the required flexibility to accommodate multiple demanding mobility profiles, avoiding the \emph{one size fits all} approach \\ \hline
         
    \end{tabular}
    \label{tab:my_label}
\end{table*}


	\subsection{\textcolor{red}{Centralized vs. Hierarchical vs. Distributed Solution}}
	While a centralized solution might offer optimality given its global view, a distributed approach can offer more reliability by eliminating the Single Point of Failure (SPoF) problem as well as avoiding congestion at a specific network node. Instead, a hierarchical approach can incorporate the benefits of both aforesaid techniques. For example, in LTE, MME is the mobility management entity with the Serving Gateway (S-GW) being the mobility anchor, and hence, it is a centralized solution. \textcolor{red}{However, Distributed Mobility Management (DMM) \cite{Liu2015} assists in decentralization of the traditional MM mechanisms, wherein instead of having a single MM anchor for all the flows on a UE, the anchors are now distributed. By distribution of MM anchors here we mean that, when a flow is initiated to/from a UE, the anchor may be chosen dependent on the flow requirements. For example, given a new flow originating to/from a UE, a MM anchor is chosen which might be very close to the UE to assist in network offloading purposes, whereas pre-existing flows might still be served from the MM anchors to which they were first assigned, so as to avoid service disruptions. }
	Hence, it would provide more reliability. The hierarchical method on the other hand, will combine the centralized and distributed approaches to offer the reliability of the distributed approach (through decentralization of mobility anchors) and the optimality of the centralized approach (e.g. through master and slave network management entities). An example of such a distributed/hierarchical approach can be found in the upcoming 5G networks, wherein through SDN and NFV there is a separation between the CP, i.e., Access and Mobility Function (AMF)- Session Management Function (SMF) for mobility management, and the data plane (DP), i.e., OpenFlow (OF) switches, etc., \cite{Yang2016, Chen2016}.
	\subsection{\textcolor{red}{Computational Resources}}
	The computational resou\-rce locations and their corresponding computational po\-wer will determine the degree to which the mobility management mechanisms can be distributed. For example, edge clouds can aid not only in MM related computation (e.g., RAT and AP selection) but can also enable faster access to content through caching. In addition, for 5G and beyond networks, it will also be critical for the MM mechanisms to determine whether services need to be migrated or replicated \cite{Urgaonkar2015, Wang2018,Frangoudis2018}, so as to maintain service continuity and hence ensure the required QoS. Note that, by service replication we mean that the services being requested by a user undergoing a mobility event are replicated to other edge servers. Further, by service migration \cite{Urgaonkar2015, Machen2018} we imply that the services being accessed by a user undergoing a mobility event are migrated to the next edge cloud server where the user is expected to move to.
	%
	\subsection{\textcolor{red}{Backhaul Considerations}}
	Network densification and the prohibitively expensive nature of installing optical fibre as backhaul \cite{Jaber2016b} will render the backhaul scenario in 5G and B5G wireless networks to be extremely heterogeneous, i.e., they will be composed of both wired and wireless links. Further, the backhaul wireless links will consist of multiple radio access techniques such as microwave, mmWave, VLC or LiFi, co-existing together \cite{Chowdhury2018}. These transformatory trends will need to be taken into consideration while developing future MM mechanisms, as:
	\begin{itemize}
	    \item Congestion or multiple-hops in the backhaul can impact the E2E latency, and consequently, the perceived QoS \cite{Rony2017}. 
	    \item Backhaul reliability will be critical given the relatively poor penetration capability of mmWave \cite{Bai2013} and additionally, strong atmospheric absorption features for VLC \cite{Boulogeorgos2018}. Thus, during mobility, attaching to an AP with a poor backhaul link quality can correspondingly lead to degradation in QoS since, there can be increased packet loss or even an outage altogether.
	\end{itemize}
	\subsection{\textcolor{red}{Context}}
	A multitude of parameters, such as user mobility profiles, type of flows, network and user policies, AP signal quality, network load, backhaul-fronthaul options, etc., constitute the context. Additionally, MM mechanisms for 5G and B5G networks will have to service users with different mobility profiles, accessing different services. Hence, the available contextual information will be valuable for any future MM mechanism. For example, in \cite{Andreev2014}, network load aware MM methods present an improvement of 75\% in throughput at the cell edge as compared to the context agnostic methods, thus reinforcing the aforesaid criteria. 
	\subsection{\textcolor{red}{Granularity of Service}}
	Granularity in MM services (e.g., based on flow, subscriber or mobility profile) will be an important component for MM methods to provision optimal solutions for 5G and B5G networks. Further, the type of granularity offered, i.e., per-flow based, mobility based, etc., will depend on the user context as well as the network conditions. Hence, innovative mechanisms like the Mobility Management-as-a-Service (MMaaS) paradigm \cite{Jain2017} will be required. In MMaaS, on-demand MM solutions can be employed by or assigned to UEs. For example, if a device is moving at a high speed ($\sim$ 300km/h) and there is another device, say an IoT device, that is stationary, then a mobility based granularity of service can be adopted. Based on this service granularity provision, the high mobility device can be allocated resources on macro-cells whilst the stationary device can be served by small cells. \textcolor{red}{Another important example being that of network slices. Network slicing, the concept, typically refers to a resource based logical slicing of the existing network infrastructure to support multiple verticals and corresponding operators that serve them \cite{Zanzi2018}. In such scenarios, on-demand MM will be necessitated by the network slices, as they will cater to services with differing mobility requirements and patterns, such as the URLLC and eMBB services.} 
	
	\subsection{\textcolor{red}{D2D Service Availability}}
	The availability of D2D services will determine how the mobility management mechanism is executed, as D2D can assist in providing seamless mobility through CP information and/or DP data forwarding. \textcolor{red}{This will be specially relevant in scenarios involving V2X \cite{Molina-Masegosa2017}, wherein for example, the vehicles, that are outside the coverage area of the infrastructure network (IN) or are experiencing a deep fade with the IN, can exchange data with it by relaying their information through other vehicles, over the PC5 interface \cite{Molina-Masegosa2017}, that might be nearby and within the coverage area of the IN or are experiencing better channel conditions with it.}
	\subsection{\textcolor{red}{Physical Layer Considerations}}
	The introduction of massive MIMO and mmWave technology will certainly impact current MM methods. Concretely, in urban environments the mmWave links will face extensive blockage alongside their limited range due to the propagation characteristics. Hence, this will require densification, which introduces the possibilities of frequent handovers (FHOs). Here by FHOs, we refer to the fact that in a dense network environment, such as those in 5G, the users will be subjected to handover (HO) scenarios more frequently as compared to that in the current networks.  On the other hand, beamforming through massive MIMO antennas can be utilized to track moving users and hence, provide them with high QoS through higher throughput and better localization services. 
	
	Further, for B5G networks, VLC and meta-surfaces have emerged as the main enablers. Note that, VLC will be challenged extremely by the existing environment. This is so because, it operates in the Terahertz range of frequencies, thus making most objects in the environment as blocking agents. Also,  meta-surfaces will lead to programmable environments, which will create the issue of dimensionality for an optimal solution. 
	
	Henceforth, the physical (PHY) layer techniques require consideration in any MM mechanism development for 5G and beyond networks.
	\subsection{\textcolor{red}{Control Plane Signaling}}
	An important target of future MM mechanisms will be to reduce the CP signaling induced during handovers. Studies such as \cite{Jain}, have proposed enhanced handover signaling mechanisms for an SDN-based core network architecture, such that the transmission and processing cost as well as the overall latency during a handover process is reduced whilst ensuring the Capital Expenditure (CAPEX) does not rise significantly. Such a procedure will enhance the QoS for the user while switching access points and hence, will be critical to the future MM suite.  \newline

\noindent \textcolor{red}{Although, a complete overhaul of MM mechanisms for future wireless networks might result in optimal solutions, the time to develop and market them will be correspondingly longer. Hence, in the following sections, we perform a novel qualitative analysis for the various legacy as well as current state-of-the-art mechanisms and standardization efforts, and evaluate their suitability as \emph{enablers for MM} in 5G and beyond wireless networks.} 


\section{\textcolor{red}{Qualitative Analysis Criteria}}

 Present day MM mechanisms and standards are extremely stable and also readily implementable. Given the challenging nature of 5G and B5G network scenarios, it is of significant interest that these mechanisms and standards be explored for their potential inclusion -- whole or in part -- as enablers for future MM solutions. Hence, we perform a novel qualitative analysis of these mechanisms on the basis of reliability, flexibility and scalability: the three pillars of any future MM strategy.
 \begin{sidewaystable*}
\renewcommand{\arraystretch}{1.2}
\caption{Governing Parameters for the Reliability, Scalability and Flexibility of a MM mechanism/standard}
\centering
\begin{tabular}{|p{0.7 cm}|p{3.8cm}|p{2.5cm}|p{0.7 cm}|p{3.5cm}|p{2.5cm}|p{0.7 cm}|p{3.5cm}|p{2.5cm}|}
\hline
\textbf{\#} & \textbf{Reliability} & \textbf{Contribution to Reqs.} & \textbf{\#} & \textbf{Flexibility} & \textbf{Contribution to Reqs.} & \textbf{\#} & \textbf{Scalability} & \textbf{Contribution to Reqs.} \\ \hline
\textcolor{red}{RL1.} & Redundancy in the number of flows, connections, etc. & \textcolor{red}{R7} &\textcolor{red}{FL1.} & Granularity of service. E.g. per flow, per connection, per user, etc.& \textcolor{red}{R9, R11} & \textcolor{red}{SL1.} &Manageable number of connections with increasing number of users & \textcolor{red}{R3, R9} \\ \hline
\textcolor{red}{RL2.} & Seamless handover capability$^\dagger$ & \textcolor{red}{R1, R7, R8} &\textcolor{red}{FL2.} & Capability to enable connectivity to multiple APs & \textcolor{red}{R1, R9} &\textcolor{red}{SL2.} &Manageable signaling load with increasing number of users & \textcolor{red}{R3, R9}\\ \hline
\textcolor{red}{RL3.} & Decentralization & \textcolor{red}{R3, R4} &\textcolor{red}{FL3.} &Handover service support at multiple network levels. E.g. Core network, Access network, etc. & \textcolor{red}{R4, R9} &\textcolor{red}{SL3.} & Manageable processing load with increasing number of users/devices & \textcolor{red}{R3, R9}\\ \hline
\textcolor{red}{RL4.} & Fast path re-routing at CN & \textcolor{red}{R5, R10} &\textcolor{red}{FL4.} & Handover decision making utilizing multiple parameters. E.g. network load, requested QoS, etc.& \textcolor{red}{R1, R9} &\textcolor{red}{SL4.} & Decentralization & \textcolor{red}{R4}\\ \hline
\textcolor{red}{RL5.} & Congestion aware & \textcolor{red}{R2} &\textcolor{red}{FL5.} &Context awareness & \textcolor{red}{R2, R9, R10} &\textcolor{red}{SL5.} &Ease of implementation and integration & \textcolor{red}{R6}\\ \hline 
\multicolumn{9}{l}{\textcolor{red}{$^\dagger$Here seamless handover capability refers to the ability of a MM mechanism to permit vertical (inter-RAT) as well as horizontal (intra-RAT) handover.}}
\end{tabular}
\end{sidewaystable*} 
\textcolor{red}{As part of this qualitative  analysis, we firstly present a detailed description of these three criteria, as follows}: 
\begin{itemize}
    \item[1.] \emph{Reliability} will help to determine whether the MM mechanisms employed will be able to ensure guaranteed and continuous service in any given network topology. Such reliability requirements entail not only continuous connectivity whilst traversing a geographic area, they also include reliability in delivery of packets for critical and delay sensitive services. Further, reliability from a MM mechanism also envelops factors such as tolerance to congestion (through for example, Distributed MM), ensuring faster yet trustworthy re-connection and authentication whilst mobile, ensuring appropriate levels of redundancy in the number of flows, connections, and hosts, and also ensuring appropriate resource allocation for users with myriad mobility and application profiles at the edge, access and core network. 
   
    \item[2.] \emph{Flexibility} as a qualitative analysis tool helps to determine the adaptability that MM mechanisms will provide to the network, which as discussed will be heterogeneous and dense in all perceivable aspects. The flexibility provisioned by MM mechanisms for future networks hence envelops factors such as the ability to formulate and deploy MM policies depending on individual user profiles, flow profiles or based on a slice profile. Further, ensuring the possibility of multi-connectivity through various layers such as transport layer (Stream Control Transmission Protocol (SCTP)/ Multi-Path Transmission Control Protocol (MPTCP)), IP layer (Multi-homing), Medium Access Control (MAC)-PHY layer (Dual Connectivity), will be an important factor for ensuring a flexible MM policy. Additionally, factors such as multi-objective access point selection/user association taking into account factors such as congestion, QoS requirements, backhaul reliability, etc., will be critical to a flexible MM mechanism.
    
    \item[3.] \emph{Scalability} aspect allows one to determine if the future MM mechanisms can serve the increasing number of user devices with a corresponding increase in requested QoS with heterogeneous mobility profiles. A measure of scalability of MM mechanisms can be gained by analyzing factors such as number of connections that can be managed given an increasing number of user devices, management of the signaling load generated due to mobility events, management of the increasing load due to processing the many CP messages generated in mobility events, as well as the ability to permit decentralization (which in essence would ensure scalability) and being easily deployable on a large scale given a new MM mechanism.  
    
\end{itemize}

\textcolor{red}{We summarize the aforesaid criteria into a list of parameters for each criteria and present them in Table 2. Additionally, we also indicate the requirements (Table 1) for whose fulfilment each of these parameters contribute towards.  Note that, compliance with each of the stated parameters in Table 2 for the reliability, flexibility and scalability criteria will be essential towards ensuring that the MM mechanism under consideration satisfies the requirements (Table 1) defined for the upcoming 5G and beyond networks. We now elaborate upon the parameter-requirement relationships that have been illustrated in Table 2, with the objective of enhancing the comprehensiveness of the evaluation criteria. } 

\subsection{\textcolor{red}{Reliability: Parameter to Requirement mapping}}
\textcolor{red}{The provision of redundancy in the number of flows and connections, i.e., by satisfying parameter \textit{RL1}, can help fulfil requirement \textit{R7} presented in Table 1. This is so because, redundancy in connections will help overcome the fragile nature of wireless channels in the frequency bands that constitute VLC and mmWave communications. Next, satisfying the parameter \textit{RL2} will contribute towards fulfilling the requirements \textit{R1}, \textit{R7}, and \textit{R8} (Table 1). Here, the ability to provision seamless handover assists in supporting mobility amongst multiple RAT(s) (\textit{R1}), supporting multi-connectivity and thus reliability (\textit{R7}), and utilize enhanced localization capabilities to accomplish the same in dense urban scenarios (\textit{R8}). Additionally, the \textit{RL3} parameter for the reliability criteria, when satisfied, will help to fulfill the \textit{R3} and \textit{R4} requirements (Table 1). The reason being, decentralization will allow for efficient handling of the number of devices (\textit{R3}). Moreover, to establish an effective level of decentralization, such as for accessing cached data at the edge and in the IMS core, enablers such as NFV and Mobile Edge Computing (MEC) will be utilized (\textit{R4}). Furthermore, the \textit{RL4} parameter holds significant relevance towards fulfilling the requirements \textit{R5} and \textit{R10} (Table 1). Specifically, fast path re-routing in the CN ensures that the increased dynamism, due to the mobility of both the UE and APs (\textit{R5}), is catered to in the CN. In addition, data path modifications, due to service migration and service replications, do not lead to extensive delays is also ensured through parameter \textit{RL4}. Lastly, satisfying the \textit{RL5} parameter will help towards fulfilling the \textit{R2} requirement (Table 1), since guaranteeing congestion awareness helps service the different QoS requirements of the applications, such as virtual reality and emergency services, with better reliability.}

\subsection{\textcolor{red}{Flexibility: Parameter to Requirement mapping}}
\textcolor{red}{When a MM mechanism under study satisfies the flexibility parameter \textit{FL1}, it correspondingly helps to fulfil the \textit{R9} and \textit{R11} requirements (Table 1). This is so because, \textit{FL1} states that a MM mechanism should support granularity of service. This will correspondingly assist in accommodating the multitude of service requirements independently (\textit{R9}) as well as avoid the \textit{one size fits all} approach (\textit{R11}). Next, \textit{FL2} parameter will help in satisfying the \textit{R1} and \textit{R9} requirements (Table 1). Essentially, the capability to be able to connect with multiple APs will assist in multi-RAT MM (\textit{R1}) as well as in provisioning enhanced agility for MM mechanisms in a dense and heterogeneous network (\textit{R9}). Further, when the \textit{FL3} parameter is satisfied, it helps to fulfil the \textit{R4} and \textit{R9} requirements. The reason being, to enable handover support at multiple levels of the network, usage of SDN, NFV and MEC platform will be necessitated for efficient implementation (\textit{R4}). Moreover, such multi-level handover support will also provision flexibility for the network (\textit{R9}). Additionally, satisfying parameter \textit{FL4} enables the MM mechanism under study to contribute towards satisfying the \textit{R1} and \textit{R9} requirements (Table 1). Specifically, having a handover decision mechanism that utilizes multiple parameters aids in handling MM amongst multiple RAT(s) more flexibly and hence, efficiently (\textit{R1}). Also, such strategies will ensure that alongside flexibility, solutions are computationally tractable and energy efficient (\textit{R9}). Finally, parameter \textit{FL5}, when satisfied, will be relevant for the fulfilment of requirements \textit{R2}, \textit{R9} and \textit{R10} (Table 1). To elaborate, the context awareness feature of a MM mechanism will assist in provisioning MM support dependent on application, user and network context (\textit{R2}), flexibility to handle the increased heterogeneity in the network (\textit{R9}), and ensure QoS whilst performing complex tasks such as migrating or relocating services based on user mobility events (\textit{R10}) through appropriate path and resource management.}

\subsection{\textcolor{red}{Scalability: Parameter to Requirement mapping}}
\textcolor{red}{For the scalability criteria, when parameter \textit{SL1}, \textit{SL2} and \textit{SL3} are satisfied by a MM mechanism, they correspondingly also assist in fulfilling the \textit{R3} and \textit{R9} requirements (Table 1). Concretely, the ability to be able to manage increasing number of connections, signaling load and processing load with the number of increasing users will correspondingly assist in handling a user density of more than $10^6$ devices per $km^2$ in 5G and beyond networks (\textit{R3}). Also, they will help in ensuring the required scalability to accommodate the increasing heterogeneity in the network as well as the corresponding tractability of the MM solution (\textit{R9}). Next, when parameter \textit{SL4} for the scalability criterion is met, it helps to fulfil the \textit{R4} requirement (Table 1). Specifically, to accomplish decentralization objective the MM mechanism under study will need to utilize enablers such as NFV and MEC. Lastly, satisfying parameter \textit{SL5} will help to meet the requirement \textit{R6} (Table 1). The reason being that, ease of implementation usually arises from the fact that a MM mechanism has been used/deployed before, as well as is suitable to accommodate legacy devices whilst catering to a new set of service and devices. Hence, satisfying the \textit{SL5} parameter will assist in ensuring that backwards compatibility requirements (\textit{R6}) are adhered to.\\}

 \noindent \textcolor{red}{And so, from the aforementioned elaborate understanding of the mapping, it can be deduced that the criteria chosen for our qualitative analysis are comprehensive in nature and approach.} Moreover, and considering only the 5G networks since their KPIs have been defined \cite{Elayoubi2016}, provisioning beyond 99.999\% reliability will be ensured through the reliability metric during mobility scenarios. Further, latency less than 5 ms for connected cars and 10 ms for virtual reality and broadband applications, will be guaranteed through the reliability and flexibility metric. \textcolor{red}{Specifically,} the reliability metric will help provision congestion awareness, reliable link selection, etc., while flexibility will allow multiple type and number of connections during mobility scenarios. In addition, support for nearly 1 million devices per km$^{2}$ with different application and mobility profiles will be ensured through the scalability criterion. \textcolor{red}{Consequently, this further reinforces the comprehensiveness of the criteria chosen for the  qualitative analysis that follows.}

\section{\textcolor{red}{Legacy mechanisms and standards: 5G and Beyond MM enablers?}}

\textcolor{red}{We evaluate certain widely employed/studied legacy standards and mechanisms based on the criteria (reliability, flexibility and scalability) listed in Table 2. It is important to state here that, the goal of the following analysis is not to compare the considered standards and mechanisms against each other but rather to highlight the extent of their suitability for 5G and beyond networks.}

\subsection{IETF MPTCP-SCTP}

\subsubsection{\textcolor{red}{Discussion}}
\textcolor{red}{Being transport layer protocols, MPTCP (through multiple TCP connections) \cite{Ford2011, Ford2013} and SCTP (through its multi-homing capabilities) \cite{Stewart2007} can provide multiple TCP paths for flows originating at the host. Generally utilized for increasing data rates \cite{Klein2011} and improving the QoS, the provision of multipath redundancy \cite{Zannettou2016, Phung2019, Liu2017a, Natarajan2009} and congestion awareness (at the transport layer level) \cite{Raiciu2011,Wischik2011, Ignaciuk2018, Stewart2007} will facilitate reliability for 5G and beyond MM mechanisms. Additionally, MPTCP and SCTP satisfy the granularity of service criterion (by provision of per-flow level granularity of service), which will be essential for the future MM mechanisms. Further, according to \cite{Ford2011, Ford2013, wei-mptcp-proxy-mechanism-02}, for MPTCP to be implemented without altering the legacy systems, proxy servers supporting MPTCP will need to be installed in front of the legacy devices, such as the middleboxes installed by service providers. The legacy systems can then communicate with the proxies using the legacy TCP protocol, while the proxies utilize MPTCP for communicating with the destination MPTCP capable device. However, it is the requirement of these additional proxies that will impact the scalability of the MPTCP solution for 5G and beyond MM mechanisms. Moreover, for SCTP, both the user and server protocol stacks need to be updated \cite{Stewart2007}. Given the number of users in future networks, it will pose a scalability challenge for the deployment of SCTP as part of the 5G and beyond MM mechanisms.}    

\subsubsection{\textcolor{red}{Analysis}}

\textcolor{red}{Given our objective of determining the suitability of MPTCP and SCTP for 5G and beyond MM mechanisms, we enlist their \textit{pros} and \textit{cons} as follows:}

\begin{itemize}
    \item \textcolor{red}{MPTCP Pros}
    \begin{itemize}
        \item \textcolor{red}{Allows for multiple data flows at the transport layer level \cite{Ford2011, Ford2013, Liu2017a}, and hence, provisions for resiliency against connection failures, given the multipath feature \cite{Zannettou2016, Phung2019, Liu2017a}}
        \item \textcolor{red}{Provisions congestion awareness, with studies such as \cite{Raiciu2011} proposing specific congestion control methods for MPTCP}
        \item \textcolor{red}{Through its ability to divide a connection into multiple sub-flows, MPTCP provisions the capability to handle each flow independently \cite{Ignaciuk2018, Liu2017a}}
    \end{itemize}
    \item \textcolor{red}{MPTCP Cons}
    \begin{itemize}
        \item \textcolor{red}{The middleboxes installed by service providers are not optimized to support MPTCP \cite{Ford2011, Ford2013}}
        \item \textcolor{red}{MPTCP requires proxies to allow MPTCP enabled devices to take its full benefits \cite{wei-mptcp-proxy-mechanism-02}}
    \end{itemize}
    \item \textcolor{red}{SCTP Pros}
    \begin{itemize}
        \item \textcolor{red}{Allows for multiple data flows at the transport layer level \cite{Stewart2007, Natarajan2009}, and hence, provisions for resiliency against connection failures, given the multipath feature. }
        \item \textcolor{red}{Provisions congestion awareness, wherein reference \cite{Stewart2007} establishes the presence of congestion avoidance methods within the SCTP suite}
        \item \textcolor{red}{Assists in network level fault tolerance through support for multi-homing \cite{Stewart2007, Natarajan2009}}
    \end{itemize}
    \item \textcolor{red}{SCTP Cons}
    \begin{itemize}
        \item \textcolor{red}{Requires both host and destination device protocols stacks to be updated with the SCTP protocol \cite{Stewart2007}}
    \end{itemize}
\end{itemize}

\noindent \textcolor{red}{From the \textit{pros} and \textit{cons} of both MPTCP and SCTP, as listed above, it can be concretely stated that the IETF MPTCP-SCTP methods satisfy parameters \textit{RL1} (allowing for multiple flows over the network for any given user) and \textit{RL5} (provisioning congestion awareness as part of the transport layer characteristic for MM) for the reliability criterion. Further, for flexibility, from our discussion above, it is clear that IETF MPTCP-SCTP only satisfies parameter \textit{FL1} (by allowing for multiple flows, flow level granularity can be induced).} 

\subsection{IEEE 802.21}

\subsubsection{\textcolor{red}{Discussion}}
\textcolor{red}{Network layer protocols will play a critical role in ensuring seamless mobility during inter-RAT mobility events, given the fact that a change in IP anchors/addresses invariably leads to a dropped session. A significant effort in this direction is provided by IEEE 802.21, which is an inter-RAT handover protocol allowing devices to move seamlessly between the various IEEE 802.x technologies \cite{Ferretti2016, DeLaOliva2008, Eastwood2008, IEEE2014, Leon2010}. Sitting just above the MAC layer, it provides information and command service to higher layers thus permitting the users to perform a media independent handover. 3GPP technologies can also utilize this information and hence, allow devices to handover from 3GPP to IEEE 802.x RATs and vice versa. Consequently, IEEE 802.21 can provision certain degree of reliability and flexibility for 5G and beyond MM mechanisms. However, note that the protocol stack of all the users would have to be modified to implement the IEEE 802.21 mechanism.} 


\subsubsection{\textcolor{red}{Analysis}}
\textcolor{red}{For the purpose of analysis, we list the \textit{pros} and \textit{cons} of the IEEE 802.21 mechanism towards 5G and beyond MM strategies, as follows:}

\begin{itemize}
    \item \textcolor{red}{IEEE 802.21 Pros}
    \begin{itemize}
        \item \textcolor{red}{Provisions seamless handover capability, as it allows users to switch between multiple RATs \cite{Ferretti2016, DeLaOliva2008, Leon2010}}
        \item \textcolor{red}{Provisions the possibility for a UE to connect to multiple APs \cite{Ferretti2016,Eastwood2008}}
    \end{itemize}
    \item \textcolor{red}{IEEE 802.21 Cons}
    \begin{itemize}
        \item \textcolor{red}{Requires the protocol stacks of both the host and destination devices to be modified, so as to enable the IEEE 802.21 functionality \cite{DeLaOliva2008, IEEE2014}}
    \end{itemize}
\end{itemize}

\noindent \textcolor{red}{And so, given the aforesaid \textit{pros} and \textit{cons} with regards to IEEE 802.21, it can be deduced that it satisfies parameter \textit{RL2} for reliability (allowing for seamless movement between different RATs) and \textit{FL2} for flexibility (allowing for the possibility to connect with multiple RATs) criteria.}





\subsection{IETF PMIPv6}

\subsubsection{\textcolor{red}{Discussion}} 
\textcolor{red}{Proxy Mobile IPv6 (PMIPv6) is a layer-3 MM protocol that allows a network based MM solution by utilizing gateways and anchors, i.e., Mobile Access Gateway (MAG) and Local Mobility Anchor (LMA), respectively \cite{Gundavelli2008, Bernados2016}. An LMA manages multiple MAGs, and is responsible for the assigment of the IP prefix which the UE retains during its entire duration within an LMA, i.e., a PMIPv6, domain \cite{Gundavelli2008, Bernados2016}. Concretely, it is the topological anchor for the UE. On the other hand, MAG is responsible for performing mobility related signaling, on behalf of the UE, with the LMA. Furthermore, it mains the assigned IPv6 prefix as the UE roams around the MAGs within an LMA domain\cite{Gundavelli2008, Bernados2016}. It is noteworthy that PMIPv6 has also been adopted by 3GPP networks \cite{3GPP2010}, thus reflecting the maturity and reliability of the solution with regards to its utility for future MM solutions.}  
\textcolor{red}{However, being centralized in nature, it can impact the network scalability and reliability in dense and heterogeneous future network environments, as a large volume of the traffic will pass through a single anchor. This can consequently lead to SPoF and congestion \cite{Nguyen2008}, thus making it less favorable for 5G and beyond MM mechanisms. And so, certain studies such as \cite{Giust2015, Nguyen2008} provide discussions on scalable methods for PMIPv6. Specifically, in \cite{Giust2015} a PMIPv6 based DMM approach has been proposed. The DMM approach essentially aids in improving the reliability and scalability aspects, as it would provide a decentralized method (without any mobility anchors) and eliminate SPoFs. Furthermore, in \cite{Nguyen2008}, a cluster based approach was proposed to enhance the scalability of the existing PMIPv6 protocol.}

\subsubsection{\textcolor{red}{Analysis}}
\textcolor{red}{Based on the discussions carried out in Section 3.3.1, we now enlist the \textit{pros} and \textit{cons} of the PMIPv6 strategy with regards to its utility for 5G and beyond MM mechanisms, as follows:}

\begin{itemize}
    \item \textcolor{red}{PMIPv6 Pros}
    \begin{itemize}
        \item \textcolor{red}{Given that PMIPv6 is adopted by 3GPP and it forms a relatively agnostic setup for an UE towards its mobility signaling, it can thus provision seamless mobility \cite{3GPP2010, Gundavelli2008, Bernados2016}}
        \item \textcolor{red}{Through the DMM based PMIPv6 approach, decentralization can be introduced \cite{Giust2015}. Furthermore, other approaches, such as the clustering based approach in \cite{Nguyen2008}, can grant enhanced scalability and reliability to the PMIPv6 approach}
        \item \textcolor{red}{Given that it has already been adopted by 3GPP for LTE, the available implementational expertise will enhance the ease with which it can be adopted in future networks\\ \newline}
    \end{itemize}
    \item \textcolor{red}{PMIPv6 Cons}
    \begin{itemize}
        \item \textcolor{red}{In its original flavor, PMIPv6 suffers from scalability and reliability issues due to the SPoF formed by the LMA in its architecture \cite{Nguyen2008}}
        \item \textcolor{red}{An explicit treatment of PMIPv6 with regards to the parameters for flexibility criterion is missing in \cite{3GPP2010, Gundavelli2008, Bernados2016, Giust2015, Nguyen2008}  }
    \end{itemize}
    
\end{itemize}

\textcolor{red}{And so, it can be deduced that the IETF PMIPv6 in its original flavor, given its maturity in development and deployment, satisfies the seamless handover parameter \textit{RL2} in the reliability criteria. However, with enhancements from the use of DMM and cluster based methods, PMIPv6 can be decentralized and scaled thus satisfying parameters \textit{RL3} and \textit{SL4} in reliability and scalability, respectively. Furthermore, since it has already been explored and implemented in the LTE networks, it satisfies parameter \textit{SL5} owing to its relative ease of implementation as against any other new protocol.}

\subsection{LTE MM mechanisms}

\subsubsection{\textcolor{red}{Discussion}}
\begin{itemize}
    \item[\textcolor{red}{A.}]\textcolor{red}{\emph{Handover:}} \textcolor{red}{Whilst LTE mobility management derives its characteristics from the PMIPv6 MM strategy \cite{Sanchez2016}, LTE-X2 offers a method to decentralize it. In the presence of an X2 interface between two LTE eNodeB's (eNBs), instead of involving the core network for resource negotiation and data forwarding tasks, the eNBs communicate amongst themselves. This allows for a fast handover and also reduces signaling in the core network \cite{Universal2015}. And so, due to the ability of LTE-X2 to provision seamless handover alongside decentralization, it can grant reliability and scalability to 5G and beyond MM mechanisms. Further, since it provisions decentralization and reduces CN signaling, it also reduces the processing load for the CN. Hence, LTE-X2 can facilitate scalability for 5G and beyond MM. Lastly, since LTE-X2 only enables multi-level HO service support, i.e., HO can be executed either at the access (through X2 HO) or core network level (through S1 HO), it offers limited flexibility.}

    \setlength\parindent{8pt}\textcolor{red}{However, note that, LTE-S1 handover involves resource negotiation and routing decisions through the MME \cite{Universal2015}. Due to this centralized approach, there will be extensive CN signaling, which will lead to congestion and a SPoF. Thus, in its own capacity, LTE-S1 handover is not foreseen as an enabler for future MM strategies.\\} 
    
    \item[\textcolor{red}{B.}] \textcolor{red}{\emph{Traffic Offloading:}} \textcolor{red}{3GPP, through Release-10, introduced Local IP Access (LIPA) and Selected IP Traffic Offloading (SIPTO) \cite{Sankaran2012} protocols. Concretely, LIPA allows for a local breakout, wherein a mobile device can communicate with another device through a private network, i.e., the data flow does not pass through the 3GPP CN, or to a public network, if the private network connects to it \cite{Sankaran2012}. An important challenge of LIPA with regards to MM is that, session continuity for LIPA connections during mobility events is not supported.}

    \setlength\parindent{8pt}\textcolor{red}{On the other hand, SIPTO is an orthodox traffic offloading mechanism, wherein the goal is to offload the IP traffic to an eNB or a gateway that is closer to the UE. Next, during 3GPP Release-10, the concept of IP Flow Mobility (IFOM) was also introduced. IFOM allows a UE to offload, if possible, data sessions to the Wi-Fi network from the 3GPP network. Consequently, through IFOM, a UE can maintain data flows belonging to the same packet data network (PDN) connection simultaneously on both the 3GPP and the Wi-Fi network \cite{Sankaran2012}.}

    \setlength\parindent{8pt}\textcolor{red}{Given these aforesaid traffic offloading strategies, they can consequently aid in managing any increase in traffic load within the network, as well as the processing load on specific network nodes, due to the increase in the number of users/devices. Thus, these mechanisms can provision scalability for 5G and beyond MM.\\}

    \item[\textcolor{red}{C.}] \textcolor{red}{\emph{Dual Connectivity and LTE-WiFi Aggregation:} The Dual Connectivity (DC) concept allows a user to camp on two APs simultaneously. Concretely, a UE can be connected to a Small-cell (SC) and a Macro-cell (MC) at the same time, wherein the MC and SC are connected to each other via the X2 interface, and the MC is the master eNB. According to 3GPP, all control plane communications, including resource allocation on SC, are performed via the corresponding MC, i.e., the master eNB, to which the UE is associated to. Note that, DC was introduced by 3GPP for LTE during Release-12. But, it is in Release-13 that this concept matured, wherein multiple usage scenarios, architecture and the operational characteristics were defined. A detailed description of the same has been presented in \cite{Carrier2015}. Furthermore, during Release-13, the concept of LTE-WiFi aggregation (LWA) was standardized \cite{Terrestrial2013}. Through LWA, a UE can simultaneously receive packets over both the LTE and the Wi-Fi interfaces, wherein the aggregation of these two physically distinct data streams takes place at the Packet Data Convergence Protocol (PDCP) layer in the protocol stack. However, note that the LWA functionality is defined only for the downlink \cite{IbarraBarreno2017}. Henceforth, given that the DC and LWA strategies provision the ability to connect to multiple APs at the same time, they can provision reliability and flexibility for 5G and beyond MM mechanisms. \\}
    
    \end{itemize}

\subsubsection{\textcolor{red}{Analysis}}
\textcolor{red}{For the 3GPP based MM mechanisms, we firstly highlight the \textit{pros} and \textit{cons} for the handover, traffic offloading and DC and LWA strategies, as follows:}

\begin{itemize}
    \item \textcolor{red}{LTE Handover Pros} 
    \begin{itemize}
        \item \textcolor{red}{The LTE-X2 and S1 mechanisms together offer handover support at the access and core network level \cite{Universal2015}}
        \item \textcolor{red}{Through LTE-X2 handover mechanism, CN signaling can be avoided \cite{Universal2015}}
        \item \textcolor{red}{LTE-X2 permits decision making for a handover to be taken at the access network level. Hence, it reduces the processing load on the CN entities as well and also permits fast handover capabilities \cite{Oh2014, Universal2015}}
    \end{itemize}
    \item \textcolor{red}{LTE Handover Cons}
    \begin{itemize}
        \item \textcolor{red}{The S1 based handover mechanism involves signaling through the CN, which creates increased load on the CN  \cite{Oh2014} as well as introduces SPoFs}
    \end{itemize}
    \item \textcolor{red}{LTE Traffic Offloading Pros}
    \begin{itemize}
        \item \textcolor{red}{Provision a method for managing the traffic load given that the number of users/devices will increase significantly \cite{Sankaran2012}}
        \item \textcolor{red}{Provision a method for managing the processing load in the network nodes \cite{Sankaran2012}}
    \end{itemize}
    \item \textcolor{red}{LTE Traffic Offloading Cons}
    \begin{itemize}
        \item \textcolor{red}{LIPA does not support session continuity during mobility events, as well as it requires an additional gateway \cite{Sankaran2012}}
        \item \textcolor{red}{SIPTO is not helpful in mitigating radio congestion \cite{Sankaran2012}}
        \item \textcolor{red}{IFOM is significantly harder to implement as it necessitates coordination with the non-3GPP networks \cite{Sankaran2012}}
    \end{itemize}
    \item \textcolor{red}{LTE DC and LWA Pros}
    \begin{itemize}
        \item \textcolor{red}{Provisions the ability to connect to multiple 3GPP as well as Non-3GPP RATs \cite{Carrier2015,Terrestrial2013, IbarraBarreno2017, He2010}}
        \item \textcolor{red}{Provisions the capability to have multiple physical paths for data transmission, and thus better fault tolerance \cite{Carrier2015,Terrestrial2013, IbarraBarreno2017, He2010}}
    \end{itemize}
    \item \textcolor{red}{LTE DC and LWA Cons}
    \begin{itemize}
        \item \textcolor{red}{3GPP LWA is only applicable for downlink}
    \end{itemize}
\end{itemize}

\textcolor{red}{From the \textit{pros} and \textit{cons} for the LTE MM mechanism, it is clear that they provision redundancy in data paths (through DC and LWA), decentralization (through X2 and traffic offloading) and seamless handover (through X2 and S1 handover), thus satisfying \textit{RL1}, \textit{RL2} and \textit{RL3}  parameters for the reliability criterion. Further, for the flexibility criterion, LTE MM mechanisms offer the possibility of a multi-level HO support (through X2 and S1 handover) as well as the ability to connect to multiple APs/RATs at the same time (through DC and LWA), thus satisfying parameters \textit{FL2} and \textit{FL3} for flexibility. Lastly, LTE MM mechanisms offer enhanced support with regards to the scalability criterion for 5G and beyond MM, as they satisfy parameters \textit{SL2} to \textit{SL5}, given their decentralization, ease of integration, multi-level handover mechanisms (X2 and S1 handover), and traffic offloading characteristics.}
\subsection{Non-3GPP Multi-Connectivity Solutions}

\subsubsection{\textcolor{red}{Discussion}} 
Multi-connectivity enables the users to establish and maintain physical and logical connections to multiple access points (possibly belonging to different RAT(s)) at the same time. \textcolor{red}{Certain standards and mechanisms, apart from those developed by 3GPP (Section 4.4), that utilize this concept are ITU-Vertical multihoming (ITU-VMH) and the Co-ordinated Multipoint (CoMP) strategy.} 

\textcolor{red}{Specifically, ITU-VMH permits the user to camp on more than one RAT, via multiple physical channels, at any given moment \cite{ITU-T2011}. Through such provision, ITU-VMH ensures path redundancy. Further, through interactions between the various Open Systems Interconnection (OSI) layers, techniques such as MPTCP/SCTP in combination with ITU-VMH can also aid in the provision of path redundancy \cite{ITU-T2011}. And so, ITU-VMH via its redundancy and seamless handover capabilities ensures reliability for 5G and beyond MM mechanisms. Note that, the seamless handover capability is facilitated by the ability of ITU-VMH to allow the user to connect to a multitude of APs, thus reducing the possibility of outage (as compared to standard HO process) during mobility events. Further, via the provision of multi-connectivity, ITU-VMH also permits per-channel granularity of service. Hence, it also provisions flexibility for future MM mechanisms. However, ITU-VMH, like the IEEE 802.21 standard, would require a transformation in the protocol stack to permit efficient resource allocation at all the protocol layers \cite{ITU-T2011}. Such a transformation might be difficult to scale to all the user devices, and hence, ITU-VMH is not a very scalable solution for 5G and B5G networks.} 



\textcolor{red}{Lastly, the Co-ordinated Multipoint (CoMP) strategy involves multiple access points co-ordinating with each other to serve a given user \cite{Irmer2011}. Similar to ITU-VMH, CoMP can also provision path redundancy as well as seamless handover capability, owing to its coordinated feature. And hence, it is also a reliable strategy for future MM strategies. Further, similar to ITU-VMH, CoMP can configure multi-connectivity alongside per-channel granularity (multiple APs permit multiple channels for transmission of data and hence, per-channel granularity of service can be provisioned) \cite{Irmer2011}. Consequently, it is qualitatively a flexible mechanism for 5G and B5G networks. However, since CoMP will involve centralized scheduling operations, it will lead to SPoF as well as challenge the scalability of backhaul networks. Consequently, this also renders CoMP as not being a very scalable proposition towards the objective of developing 5G and beyond MM solutions.}     

\subsubsection{\textcolor{red}{Analysis}}
\textcolor{red}{We now present the \textit{pros} and \textit{cons} for ITU-VMH and CoMP strategies, as follows:}

\begin{itemize}
    \item \textcolor{red}{ITU-VMH Pros}
    \begin{itemize}
        \item \textcolor{red}{Provisions path redundancy through multi-homing \cite{ITU-T2011}}
        \item \textcolor{red}{Provisions the capability to connect to multiple RAT(s) at any given time \cite{ITU-T2011}}
        \item \textcolor{red}{Per-channel granularity of service is possible}
    \end{itemize}
    \item \textcolor{red}{ITU-VMH Cons}
    \begin{itemize}
        \item \textcolor{red}{It will require the transformation of the entire protocol stack \cite{ITU-T2011}}
    \end{itemize}
    \item \textcolor{red}{CoMP Pros}
    \begin{itemize}
        \item \textcolor{red}{Provisions path redundancy through its ability to coordinate data transmission from multiple APs, which may also belong to different RATs \cite{Irmer2011, Sun2018a}}
        \item \textcolor{red}{Provisions the capability to connect to multiple RAT(s) at any given time \cite{Irmer2011, Sun2018a}}
        \item \textcolor{red}{Through the use of multiple APs for transmission, per-channel granularity of service is made possible}
    \end{itemize}
    \item \textcolor{red}{CoMP Cons}
    \begin{itemize}
        \item \textcolor{red}{Centralized processing introduces the possibility of SPoF \cite{Irmer2011, Lee2012}}
        \item \textcolor{red}{Backhaul networks will need to have extremely high capacity and extremely low latency characteristics, so as to support CoMP whilst maintaining QoS \cite{Lee2012}}
    \end{itemize}
\end{itemize}

\textcolor{red}{Concretely, ITU-VMH and CoMP satisfy parameters \textit{RL1} (allowing for the possibility of redundant physical connections) and \textit{RL2} (allowing for seamless mobility) for the reliability criterion, and parameters \textit{FL1} (provisioning the possibility of per-channel granularity for MM) and \textit{FL2} (allowing for the possibility of connecting to multiple RATs/APs) for the flexibility criterion.}

\subsection{RSS based AP selection methods}

\subsubsection{\textcolor{red}{Discussion}} 
\textcolor{red}{The erstwhile Received Signal Strength (RSS) based methods employ a very simplistic approach to AP selection, i.e., comparing the detected AP link quality (RSSI/RSRP/RSRQ) levels \cite{3GPP2011,3GPP2010, Xenakis2014}. The aforesaid simplistic nature can hence permit scalability for the future MM mechanisms as it is easy to implement, and does not entail a high processing and signaling load either. However, such an approach can be plagued by multiple issues. For example, APs with a good RSS might be overloaded (as more users will be assigned to them) whilst others maybe under-utilized. Such a scenario also implies that RSS based methods are not reliable as a better RSS does not always guarantee better QoS, since, congestion will lead to degraded QoS. Moreover, in dense scenarios, even with the implementation of a hysteresis, UEs will be subject to FHOs due to the fluctuating RSS and availability of multiple candidate APs. This further exemplifies the unreliability of RSS based methods. Additionally, these methods are one-dimensional, given that they consider only RSS as a decision parameter. The RSS methods also do not provision any granularity of service, context awareness, multiple levels of HO support, etc. Hence, they do not offer any flexibility to the MM mechanisms for 5G and B5G networks.}

\begin{sidewaystable*}
\renewcommand{\arraystretch}{1.1}
\caption{\textcolor{red}{Compliance with the Reliability, Scalability and Flexibility criteria for the legacy MM mechanism/standard}}
\centering
\color{red}\begin{tabular}{|*{14}{>{\centering\arraybackslash}p{1.25 cm}|}}
\cline{3-14}
\multicolumn{2}{c|}{} & \multicolumn{12}{c|}{\textbf{Mechanisms}}\\ \cline{3-14}
\multicolumn{2}{c|}{} & \multicolumn{2}{p{2.5cm}|}{\textbf{IETF MPTCP-SCTP}} & \multicolumn{2}{p{2.5cm}|}{\textbf{IEEE 802.21}} & \multicolumn{2}{p{2.5cm}|}{\textbf{IETF PMIPv6}} & \multicolumn{2}{p{2.5cm}|}{\textbf{LTE MM mechanisms}} & \multicolumn{2}{p{2.5cm}|}{\textbf{Non-3GPP Multi-connectivity solutions}} & \multicolumn{2}{p{2.5cm}|}{\textbf{RSS based handover methods}} \\ \hline
\multicolumn{2}{|c|}{} & Cnf.$^\dagger$ & Refs.$^\delta$ & Cnf. & Refs. & Cnf. & Refs. & Cnf. & Refs. & Cnf. & Refs. & Cnf. & Refs. \\ \hline

\multirow{5}{*}{\rotatebox{90}{\textbf{Reliability}}} & \textbf{RL1} & \Checkmark & \multirow{5}{*}[-0.2em]{\cite{Klein2011, Zannettou2016, Phung2019,Liu2017a, Natarajan2009,Raiciu2011, Wischik2011}} & $\times$ & \multirow{5}{*}[-0.2em]{\cite{Ferretti2016, DeLaOliva2008}} & $\times$ & \multirow{5}{*}[-0.1em]{\cite{Gundavelli2008, Bernados2016}} & \Checkmark & \multirow{5}{*}[-0.1em]{\cite{Universal2015, He2010}}  & \Checkmark & \multirow{5}{*}[-0.2em]{\cite{ITU-T2011,Irmer2011, Sun2018a}}& $\times$ & \multirow{5}{*}[-0.1em]{\cite{3GPP2011, 3GPP2010}} \\ \cline{2-3}\cline{5-5}\cline{7-7}\cline{9-9}\cline{11-11}\cline{13-13}
& \textbf{RL2} & $\times$ & & \Checkmark & & \Checkmark &  & \Checkmark &  & \Checkmark & & \Checkmark& \\ \cline{2-3}\cline{5-5}\cline{7-7}\cline{9-9}\cline{11-11}\cline{13-13}
& \textbf{RL3} & $\times$ &  & $\times$ & & \Checkmark & & \Checkmark & & $\times$ & & $\times$ &  \\\cline{2-3}\cline{5-5}\cline{7-7}\cline{9-9}\cline{11-11}\cline{13-13}
& \textbf{RL4} & $\times$ &  & $\times$ & \cite{Leon2010} & $\times$ & \cite{Giust2015} & $\times$ & \cite{Carrier2015, Terrestrial2013, IbarraBarreno2017} & $\times$ & & $\times$ & \cite{Xenakis2014}\\ \cline{2-3}\cline{5-5}\cline{7-7}\cline{9-9}\cline{11-11}\cline{13-13}
& \textbf{RL5} & \Checkmark & & $\times$ & & $\times$ & & $\times$ & & $\times$ & & $\times$ & \\ \hline \hline 
\multirow{5}{*}{\rotatebox{90}{\textbf{Flexibility}}} & \textbf{FL1} & \Checkmark & \multirow{5}{*}[-0.1em]{\cite{Zannettou2016, Ignaciuk2018}} & $\times$ & \multirow{5}{*}[-0.1em]{\cite{Ferretti2016, Eastwood2008}} & $\times$ & \multirow{5}{*}[-0.1em]{\cite{3GPP2010, Gundavelli2008, Bernados2016, Giust2015, Nguyen2008}} & $\times$ & \multirow{5}{*}[-0.1em]{\cite{Universal2015, He2010}} & \Checkmark & \multirow{5}{*}[-0.1em]{\cite{ITU-T2011,Irmer2011, Sun2018a}} & $\times$ & \multirow{5}{*}[-0.1em]{\cite{3GPP2010, 3GPP2011}}\\ \cline{2-3}\cline{5-5}\cline{7-7}\cline{9-9}\cline{11-11}\cline{13-13}
& \textbf{FL2} & $\times$ & & \Checkmark & & $\times$ & &  \Checkmark & & \Checkmark & & $\times$ & \\ \cline{2-3}\cline{5-5}\cline{7-7}\cline{9-9}\cline{11-11}\cline{13-13}
& \textbf{FL3} & $\times$ & & $\times$ & & $\times$ & & \Checkmark &  & $\times$ & & $\times$ & \\ \cline{2-3}\cline{5-5}\cline{7-7}\cline{9-9}\cline{11-11}\cline{13-13}
& \textbf{FL4} & $\times$ & \cite{Liu2017a, Natarajan2009} & $\times$ & \cite{Leon2010} & $\times$ & & $\times$ & \cite{Carrier2015, Terrestrial2013, IbarraBarreno2017} & $\times$ &  & $\times$ & \cite{Xenakis2014,Shen2017} \\ \cline{2-3}\cline{5-5}\cline{7-7}\cline{9-9}\cline{11-11}\cline{13-13}
& \textbf{FL5} & $\times$ & & $\times$ & & $\times$ & & $\times$ & & $\times$ & & $\times$ & \\ \hline \hline
\multirow{5}{*}{\rotatebox{90}{\textbf{Scalability}}} & \textbf{SL1} & $\times$ & \multirow{5}{*}[-0.1em]{\cite{Ford2011, Ford2013}} & $\times$ & \multirow{5}{*}[-0.1em]{\cite{DeLaOliva2008, IEEE2014}} & $\times$ & \multirow{5}{*}[-0.1em]{\cite{Giust2015}} & $\times$ & \multirow{5}{*}[-0.1em]{\cite{Sankaran2012, Universal2015}} & $\times$ & \multirow{5}{*}[-0.1em]{\cite{Irmer2011, Lee2012}} & $\times$ & \multirow{5}{*}[-0.1em]{\cite{3GPP2011, 3GPP2010}} \\ \cline{2-3}\cline{5-5}\cline{7-7}\cline{9-9}\cline{11-11}\cline{13-13}
& \textbf{SL2} & $\times$ & & $\times$ & & $\times$ & & \Checkmark & & $\times$ & & \Checkmark & \\ \cline{2-3}\cline{5-5}\cline{7-7}\cline{9-9}\cline{11-11}\cline{13-13}
& \textbf{SL3} & $\times$ & & $\times$ & & $\times$ & & \Checkmark & & $\times$ & & \Checkmark & \\ \cline{2-3}\cline{5-5}\cline{7-7}\cline{9-9}\cline{11-11}\cline{13-13}
& \textbf{SL4} & $\times$ & \cite{wei-mptcp-proxy-mechanism-02}& $\times$ & & \Checkmark & \cite{Gundavelli2008, Bernados2016}& \Checkmark & & $\times$ & & $\times$ & \cite{Xenakis2014, Ahmed2014} \\ \cline{2-3}\cline{5-5}\cline{7-7}\cline{9-9}\cline{11-11}\cline{13-13}
& \textbf{SL5} & $\times$ & & $\times$ & & \Checkmark & & \Checkmark & & $\times$ & & \Checkmark & \\ \hline 
\multicolumn{14}{l}{$^{\dagger}$The conformance (Cnf.) of a given mechanism for a given criterion.} \\
\multicolumn{14}{l}{$^{\delta}$The corroborating references (Refs.), if any, for the specified conformance of a mechanism for a given criterion} 
\end{tabular}
\end{sidewaystable*}

\subsubsection{\textcolor{red}{Analysis}}
\textcolor{red}{Based on the discussion, we present here the \textit{pros} and \textit{cons} of the RSS based AP selection methods as follows: }
\begin{itemize}
    \item \textcolor{red}{RSS based methods Pros} 
    \begin{itemize}
        \item \textcolor{red}{Easy to implement, given that it has already been adopted by 3GPP \cite{3GPP2010,3GPP2011, Ahmed2014}}
        \item \textcolor{red}{Relatively low processing and signaling load, owing to its simplicity \cite{Ahmed2014}}
    \end{itemize}
    \item \textcolor{red}{RSS based methods Cons}
    \begin{itemize}
        \item \textcolor{red}{FHOs in ultra dense scenarios is a pertinent issue \cite{Shen2017}}
        \item \textcolor{red}{It is agnostic of other parameters related to the UE and the network, such as the load, UE context, etc., thus making it unreliable and one-dimensional \cite{3GPP2010, 3GPP2011, Shen2017}}
    \end{itemize}
\end{itemize}

\textcolor{red}{Given these \textit{pros} and \textit{cons}, the erstwhile RSSI based method due to its existence and maturity can ensure mobility between multiple RAT(s), hence, satisfying parameter \textit{RL2} for reliability criteria. Furthermore, owing to the aforementioned simplicity and maturity in development and deployment it also satisfies parameters \textit{SL2}, \textit{SL3} and \textit{SL5} for the scalability criterion.\\}

\noindent \textcolor{red}{To summarize, we introduce Table 3 wherein we indicate the parameters that each of the explored methods satisfies for the reliability, scalability and flexibility criteria. We also enlist the important references that have lead us to the development of Table 3, as presented in this article. From the discussions, analysis and Table 3 it can be deduced that none of the legacy mechanisms that have been studied achieve the requirements as necessitated by 5G and B5G networks. Concretely, none of the studied mechanisms satisfy all the parameters of the criteria utilized for the qualitative analysis. Notably, the 3GPP based LTE MM mechanisms provision the best basis and support for 5G and beyond MM mechanism, given that they collectively satisfy the most parameters amongst other strategies explored. }

\textcolor{red}{Additionally, through this qualitative analysis, whilst we have presented the offered capabilities from legacy mechanisms towards 5G and B5G MM, we have also exposed the gaps that exist. This reinforces the fact that a holistic MM strategy for future wireless networks still remains elusive. Hence, in the following section, we explore the current state-of-the-art in MM solutions for 5G and beyond networks.} 

\section{5G and Beyond MM: Current State of the Art}

\textcolor{red}{Global efforts have spinned up consortiums that have provided impetus to the development of 5G, including that of MM strategies.} Further, for B5G networks, such as 6G, certain collaborative efforts have already started. References \cite{Basar2019, Renzo2019, Boulogeorgos2018, Chowdhury2018, Zhao2019} highlight the advances that have been made with regards to identifying the enablers and core principles of B5G networks. Hence, in this section we first detail the current state of the art in MM mechanisms and the parameters they satisfy from Table 2. We then follow this with a first discussion in literature that elaborates upon the utility of the current state of the art in MM for B5G networks. 

\textcolor{red}{As a prologue to the aforementioned discussion, we introduce Figure 2, wherein the 5G architecture standardized by 3GPP has been presented \cite{3GPP2020}. Correspondingly, we have also presented the classification of the various mechanisms that we explore in Sections 5.1 and 5.2 with respect to the 5G architecture in Figure 2. This classification is dependent on the portion of the network that is impacted (directly or indirectly) the most by a particular MM scheme. Furthermore, we have illustrated whether the studied mechanisms are either control plane or data plane solutions. Concretely, a CP solution would primarily impact MM via either CP signaling or decisions, while a DP solution would entail provisioning alternate and more efficient data paths. A detailed discussion with regards to these classifications has been provided in Sections 5.1 and 5.2.}

\textcolor{red}{Concretely, the 5G architecture, as shown in Figure 2, consists of two main core network functions, i.e., the Session Management Function (SMF) and the Access and Mobility Management Function (AMF). The SMF communicates with the User Plane Function (UPF) over the N4 interface, while the AMF is responsible for communicating with the RAN side over the N2 interface. Furthermore, the AMF and SMF communicate with other network functions, such as the Policy Control Function (PCF), Authentication Server Function (AUSF), etc., to execute their defined functionalities within the ambit of the policies and existing user and network context. For the sake of brevity, in Figure 2 we club all of these functions into a single entity box called \textit{Network Functions}. Moreover, the AMF also has an N26 interface that connects to the Evolved Packet Core (EPC) to facilitate Inter-RAT mobility, while an N32 interface exists in the event of a change in Public Land Mobile Network (PLMN) with 5G Core (5GC) as the CN for both the visted and home networks. Note that, the interfaces \textit{N2}, \textit{N4}, \textit{N26} and \textit{N32} are all control plane paths, with the AMF, SMF and other network functions forming the control plane entities.}

\begin{figure*}
	\centering
		\includegraphics[scale=0.42]{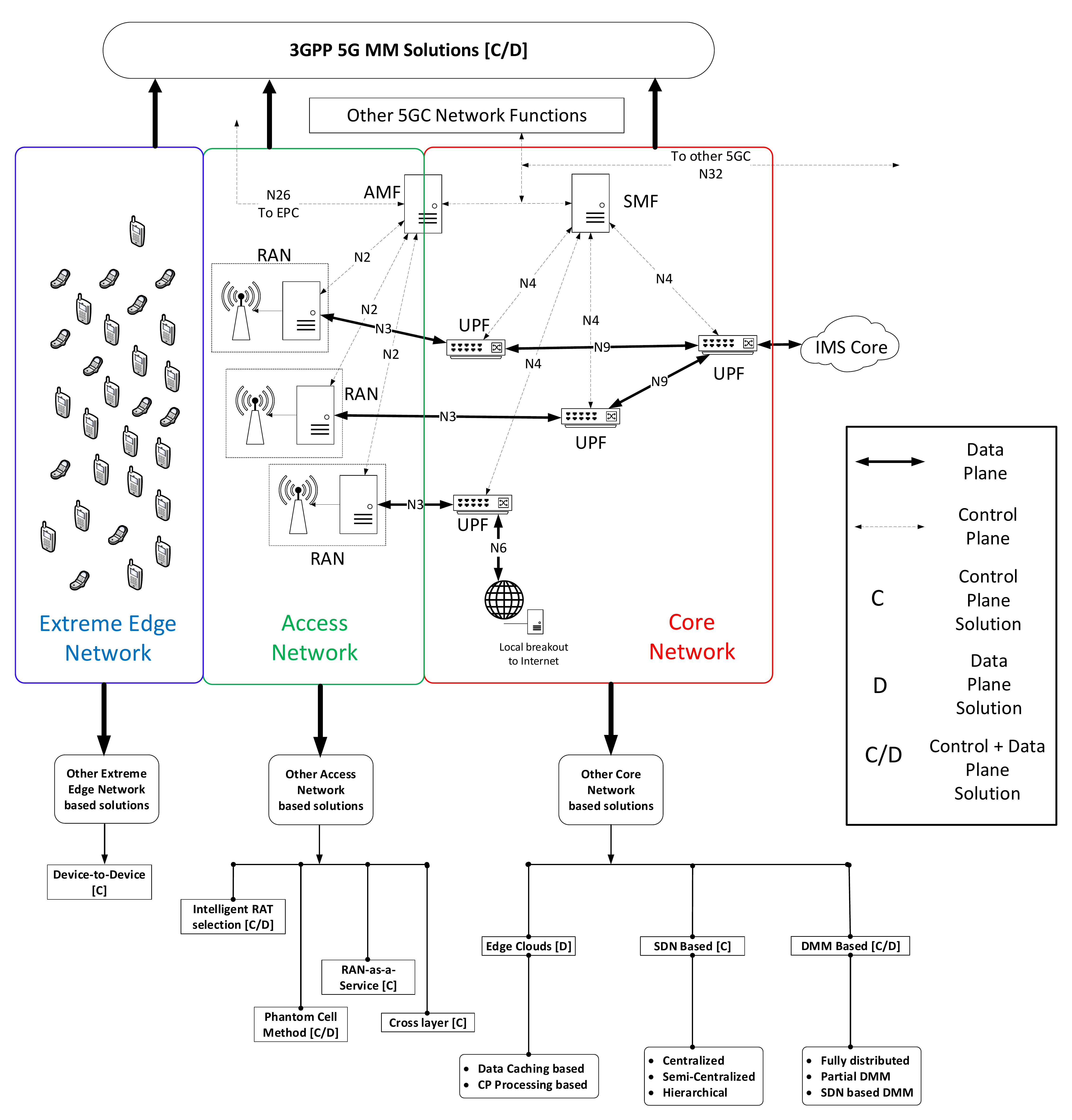}
	\caption{Classification of the state of the art in MM strategies on the 5G architecture.}
\end{figure*}

\textcolor{red}{In addition, the AMF in 5G networks is the equivalent of the MME in LTE-4G networks. It focuses on handling mobility at the access network level (such as AP selection, resource allocation, etc.). The SMF on the other hand handles the CN related tasks during mobility events (such as path re-routing, etc.). Next, in Figure 2, it can be seen that the RAN interacts with the UPF through interface N3, and the UPFs use the N9 interface to communicate amongst themselves. Also, the 5G networks provision a local breakout through the N6 interface from an UPF. The interfaces \textit{N3}, \textit{N9} and \textit{N6} constitute the data plane paths, with the RAN and UPF forming the data plane entities. Lastly, the UE, which is also a data plane entity, interacts with the AMF through the N1 interface. However, to maintain clarity, we have omitted the illustration of this interface from Figure 2. Thus, with this background, we now explore the 3GPP 5G MM mechanisms as well as other research efforts with regards to MM for 5G networks.}

\subsection{\textcolor{red}{3GPP 5G MM Mechanisms}}

\textcolor{red}{3GPP, through TS 23.501 \cite{3GPP2020}, TS 23.502 \cite{3GPP2020a} and TS 38.300 \cite{3GPP2020b}, has provided significant insights into the design and development of 5G MM strategies. New session management methods, service continuity states, UE mobility monitoring, provisioning for multi-homing, load balancing strategies, provision of on-demand MM, resource allocation due to mobility events, the new MM module, i.e., AMF, inter- and intra- next generation core (NGC) handovers, and LTE-EPC 5G-NGC interworking have been introduced in the aforesaid 3GPP specifications. These techniques through the provision of a softwarized solution and a global view of the network scenario alongside user context appear to facilitate the efficient operation of 5G and beyond MM mechanisms. Consequently, in the discussions that follow, we investigate these newly defined 3GPP MM mechanisms and elaborate their \textit{pros} and \textit{cons} for future MM mechanisms.}

\subsubsection{\textcolor{red}{Discussion}}
\begin{itemize}
    
\item[\textcolor{red}{A.}]\textcolor{red}{\textit{UE Mobility monitoring:} In TS 23.501 \cite{3GPP2020}, details with regards to how the UE mobility is monitored and the corresponding actions with regards to resource allocation and context updates have been specified. Concretely, when a UE is mobile, the 5G standards define that the AMF will be responsible for monitoring its movement and hence, its mobility pattern. Furthermore, during a UE mobility event, new resources on the destination AP are managed by the AMF through the RAT and Frequency Selection Parameter (RFSP). Such a process simplifies the identification of the required resources, as well as migration of these resources to the destination network. Moreover, the AMF manages the UE mobility event notification, i.e., it provisions details with regards to the mobility event as well as the areas of interest (Tracking areas, Cells, RANs, etc., to which a UE might migrate to). The other Network Functions (NF), such as the SMF, can subscribe to these notifications so as to employ their decisions and policies. }

\item[\textcolor{red}{B.}]\textcolor{red}{\textit{Session Management:} Through TS 23.501 \cite{3GPP2020}, the various modes that can be utilized to manage the multiple heterogeneous sessions for a given user has been defined. Notably, if a UE is connected to multiple RATs then, for a given Protocol Data Unit (PDU) session, the UE has the choice to select the access network over which this PDU session will be served. In addition, the UE, in the event of mobility or congestion, can request a PDU session to be transferred from 3GPP to non-3GPP RAT(s). Furthermore, in roaming scenarios, PDU sessions can either avail a local breakout or be routed through the home network. Specifically, each PDU session can be granted, independently, different routing modes. To do so, the SMF in the 5G CN controls and monitors the status of the data paths. Moreover, the SMF also provisions the capability of performing selective traffic routing by the application of Uplink Classifier (UL CL) on certain data plane entities, i.e., UPFs. A UPF essentially performs the function of a router in the 5G network.}

\item[\textcolor{red}{C.}]\textcolor{red}{\textit{IPv6 multihoming:} The new 5G standards, as specified in TS 23.501 \cite{3GPP2020}, have formalized the use of IPv6 multi-homing so as to reap the benefits from the multiple physical channels that will be available for use through multi-connectivity. Specifically, according to TS 23.501, more than one session anchor can be specified for a PDU session. Note that, a PDU session anchor's primary role is to assign the IPv6 prefixes that are used by the UE for a given PDU session to communicate with the public network. However, all these PDU session anchors will have a single UPF as a branching point. Next, during a mobility event, a make-before-break approach for a PDU session is adopted to provision service continuity. It must be stated here that, service continuity is ensured through the Session and Service Continuity (SSC) modes, which we will discuss next.}

\item[\textcolor{red}{D.}]\textcolor{red}{\textit{Session and Service Continuity Modes:} 3GPP, through TS 23.501 defines the SSC modes, which are critical for the network in determining the level of service continuity offered to a PDU session \cite{3GPP2020}. Concretely, three modes are defined, i.e., \textit{SSC mode 1}, \textit{SSC mode 2} and \textit{SSC mode 3}. We briefly describe them as follows:}

\begin{itemize}
    \item \textcolor{red}{\textit{SSC mode 1:} This mode ensures that the IP address is preserved. Specifically, the PDU session anchor is maintained regardless of the access technology being used by the PDU session after the mobility event. Furthermore, the IP address is maintained throughout the lifetime of the PDU session. Additionally, more PDU session anchors might be allocated for additional IP addresses, however, it is not necessary that they be maintained just like the initial IP address and PDU session anchor.}
    \item \textcolor{red}{\textit{SSC mode 2:} In this mode, if needed, the network can release a PDU session and request the UE to immediately establish a new PDU session with the same network. Moreover, if the UE has multiple PDU session anchors, the additional anchors can be released or allocated (for new IP prefixes/addresses).}
    \item \textcolor{red}{\textit{SSC mode 3:} In this mode, IP address is not preserved. This consequently makes any changes in the user plane visible to the UE. However, to ensure that an acceptable level of QoS, and hence, service continuity is maintained, a \textit{make-before-break} approach is followed. This essentially determines the destination PDU session anchor before relieving the resources the PDU session occupies at its current anchor.}
\end{itemize}

\textcolor{red}{It must be stated here that the SSC mode for a UE is selected by the SMF depending on the UE subscription details as well as the PDU session type.}

\item[\textcolor{red}{E.}]\textcolor{red}{\textit{User Plane aspects:} In 5G networks, UPFs will be utilized to handle the data plane traffic. Concretely, they can be thought of as routers, on whom the routing rules are programmed by the SMF. In TS 23.501 \cite{3GPP2020}, the aforesaid specifics have been defined. However, note that the methodology to establish these paths still involves exchanging Tunnel Endpoint Identifiers (TEIDs) between CN entities. This, as we will state in the analysis, can be a cause of increased network load. Additionally, traffic re-routing, in the event of mobility or load balancing, is handled by the SMF, wherein it sends the necessary information, such as the forwarding target information, to the UPFs. Lastly, in the event of mobility of a UE, packet buffering is also provisioned so as to minimize the loss of packets and hence, QoS. }

\item[\textcolor{red}{F.}]\textcolor{red}{\textit{Dual Connectivity:} Through TS 23.501 \cite{3GPP2020} and TS 37.340 \cite{ETSI2019}, 3GPP has also concretized and standardized the integration of Multi-RAT Dual Connectivity (MR-DC) into 5G. Concretely, the UEs will now have the capability and possibility to connect to two APs belonging to the same RAT (LTE-LTE, 5G New Radio (NR) - 5G NR) or to different RAT(s) (LTE - 5G NR). As in LTE-DC, this can be configured to allow fast-switching (fast HO), since control plane is not changed unless the Master Node is changed. Also, the UP is terminated at MC, so, no CN signalling is necessary for intra-MC HO.}

\item[\textcolor{red}{G.}]\textcolor{red}{\textit{Edge Computing:} TS 23.501 \cite{3GPP2020} defines the support for edge computing platforms in 5G networks. Concretely, these are utilized in the non-roaming or local breakout roaming modes. By local breakout, we mean that a UE can access public network without traversing the core network via additional gateways that are placed within the network. Note that, the 5G CN is responsible for selecting a UPF that is close to the UE and also has access to an edge compute node. Consequently, traffic steering is performed at this UPF towards the edge compute node.} 

\item[\textcolor{red}{H.}]\textcolor{red}{\textit{Network Slicing:} The concept of enabling a telecom operator to be able to slice its infrastructure network into logically separated networks and consequently service multiple tenants, e.g., virtual network operators, services (eMBB, URLLC, mMTC), etc., using the same, wherein the logical separation involves dynamic allocation of network resources, is termed as network slicing \cite{Zanzi2018}. 3GPP, in TS 23.501 \cite{3GPP2020}, has discussed network slicing in detail, wherein its support for roaming as well as its involvement in the inter-working process between 5G CN and LTE EPC has been elaborated. Specifically, support for migrating and translating the Single Network Slice Selection Assistance Information (S-NSSAI), which consists of the necessary information with regards to an assigned network slice for a UE, between the Home PLMN (H-PLMN) and the Visited PLMN (V-PLMN) has been detailed. Similarly, for the inter-working process, 3GPP charts out the principles for migration, translation and creation of S-NSSAIs whenever a UE undergoes mobility and changes from a 5G network to an LTE network, and vice versa. Moreover, the support has been defined for scenarios where the N26 interface, which is the standard 5G CN and LTE EPC inter-working interface, may or may not be present \cite{3GPP2020}. }

\setlength\parindent{8pt}\textcolor{red}{On the other hand, and importantly, the concept of network slicing also assists in provisioning tailor-made MM solutions for the tenants that each network slice will cater to. This consequently helps to deploy on-demand MM strategies.}

\item[\textcolor{red}{I.}]\textcolor{red}{\textit{Load Balancing and Congestion Awareness:} In TS 23.501 \cite{3GPP2020}, 3GPP has defined procedures for load balancing at the AMF and SMF, as well as congestion awareness within the core network. Concretely, two specific strategies, i.e., load balancing and load re-balancing, have been provisioned. Within the load balancing paradigm, new users incoming into an AMF region, if necessary, are directed to an appropriate AMF in order to manage the load of the AMFs. To do this, appropriate weights, indicative of the load on each AMF, are assigned and updated at appropriate intervals (typically on a monthly basis). On the other hand, if an AMF becomes overloaded, then load re-balancing is performed. Here, already registered users are migrated to other AMFs that are not overloaded while ensuring minimum service disruption \cite{3GPP2020}. Note that, the new AMF chosen should belong to the same AMF set. An AMF set is defined as the AMFs which belong to the same PLMN, have the same AMF region ID and the same AMF set ID value \cite{3GPP2020}. These parameters are pre-configured by the network operator. Lastly, 3GPP also provisions extensive details with regards to handling congestion control for the Non Access Stratum (NAS) messages. This is important from the perspective of MM, as MM messages are carried over NAS to the CN nodes. For further details with regards to the specifics of the congestion control procedures, the reader is referred to TS 23.501 \cite{3GPP2020}.}

\item[\textcolor{red}{J.}]\textcolor{red}{\textit{Cell, Beam and Network Selection:}  Through TS 23.501 \cite{3GPP2020} and in particular through TS 38.300 \cite{3GPP2020b} details with regards to cell, beam and network selection have been specified. For \textit{cell selection} these standards documents, developed by 3GPP, specify support for cell selection procedures given that the UE is in either Radio Resource Control (RRC) idle, or RRC inactive or RRC connected state. Note that, RRC idle state refers to a UE that can listen to paging channels, broadcasts and multicasts, as well as perform cell quality measurements. The RRC inactive state refers to a UE that can roam within the RAN-based notification area (RNA) without informing the NG-RAN. The RRC connected state for a UE implies that it has an active connection and data flow. Most notably, for the RRC connected state, cell mobility and beam mobility have  been specified. As the name suggests, a UE can either undergo a cell handover or it can switch between the beams that a given AP uses. To perform this, procedures for beam quality and cell quality measurements have also been defined in \cite{3GPP2020b}. The beam quality measurements are performed in the physical layer for multiple beams being transmitted by a given cell. These measurements are filtered and aggregated at the RRC layer to obtain the cell quality measurements. Note that, these quality measurements are still performed using the RSSI/RSRP/RSRQ/SINR metrics. Furthermore, in \cite{3GPP2020b} procedures for cell selection and handover involving intra- and inter-frequency handover in 5G NR, Inter-RAT handover within 5G CN, Inter-RAT handover from 5GC to EPC and vice versa, have been specified. For the sake of brevity, we do not detail these procedure and refer the reader to TS 38.300 \cite{3GPP2020b}. Moreover for Inter-RAT handovers, procedures for packet buffering and forwarding as well as data path switching, to ensure the requested QoS, have also been defined. Lastly, roaming and access restrictions are also appropriately defined based on the user subscription to both the SMF and AMF. This facilitates the selection of the right AP and PLMN for a given user \cite{3GPP2020,3GPP2020b}. }

\item[\textcolor{red}{K.}]\textcolor{red}{\textit{Inter-Working, Migration and Handover signaling:} While TS 38.300 \cite{3GPP2020b} specified certain handover procedures for both the CP and DP, a detailed description of the handover signaling, inter-working between 5G CN and EPC, and migration of PDU sessions has been provided in TS 23.502 \cite{3GPP2020a} and TS 23.501 \cite{3GPP2020}. Concretely, through \cite{3GPP2020a} the CN signaling process for the various stages in a handover, i.e., handover request, handover preparation, handover complete/cancel/reject, have been presented in detail. These handover signaling strategies have been detailed for Intra-RAT HO (N2 and Xn handovers) as well as for Inter-RAT handovers (involving 5GC and EPC). Moreover, the handover signaling procedures have also been defined for the scenarios wherein the EPC-5GC inter-working interface, i.e., N26, may or may not be present. Also note that, the 5G-N2 handover is similar to the LTE-S1 handover (specified in Section 4.4.1) and the 5G-Xn handover is similar to the LTE-X2 handover (specified in Section 4.4.1). Next, for the 5GC and EPC inter-working, in TS 23.501 \cite{3GPP2020} the principles for maintaining IP address continuity in the event of UE mobility from 5GC to EPC or vice versa have been provisioned. However, it is also specified that in the event a UE transitions from 5G to 3G or 2G and vice versa, the IP address continuity might not be maintained. Furthermore, procedures for transferring the PDN/PDU sessions established by a UE over a 4G/5G network, when it transitions to the 5GC/EPC, over the N26 interface have been provisioned in \cite{3GPP2020}. Also, traffic steering and forwarding procedures have also been elaborated. Lastly, procedures for migrating PDU sessions from non-3GPP access to the 3GPP access, when a UE undergoes a mobility event from 5GC to EPC, is also supported \cite{3GPP2020}.}

\item[\textcolor{red}{L.}]\textcolor{red}{\textit{D2D mobility support:} With the standardization of Proximity Services (ProSe) in 3GPP Release-12 and 13 \cite{Terrestrial2013}, 5G networks can utilize the capability to orchestrate data forwarding/relaying in both DP and CP. This can consequently enhance the ability of the network to provide a proactive and seamless handover procedure \cite{Jung2016}.}
\end{itemize}

\subsubsection{\textcolor{red}{Analysis}}
\textcolor{red}{Given the extensive overview with regards to the MM solutions that have been provisioned by the 5G standards \cite{3GPP2020, 3GPP2020a, 3GPP2020b}, we now, as part of our qualitative analysis, present the \textit{pros} and \textit{cons} for the same, as follows:}

\begin{itemize}
    \item \textcolor{red}{3GPP 5G MM Pros}
    \begin{itemize}
        \item  \textcolor{red}{Provisions monitoring of UE mobility, mobility event notifications and resource negotiation mechanisms at destination networks \cite{3GPP2020}}
        \item \textcolor{red}{Employs flexible session management strategies, wherein provision of per-PDU session granularity, through path selection, roaming support and traffic steering, has been detailed \cite{3GPP2020}}
        \item \textcolor{red}{Support for IPv6 multi-homing \cite{3GPP2020}}
        \item \textcolor{red}{Provision for multiple sessions and service continuity modes \cite{3GPP2020}}
        \item \textcolor{red}{Support for Multi-RAT DC \cite{3GPP2020}}
        \item \textcolor{red}{Support for Edge Computing \cite{3GPP2020}}
        \item \textcolor{red}{Network slicing information migration support in the event of inter-/intra- RAT mobility \cite{3GPP2020}}
        \item \textcolor{red}{Network slicing support for provisioning on-demand MM}
        \item \textcolor{red}{Ability to provision context awareness via network slicing}
        \item \textcolor{red}{Provision for managing core network load by introducing load balancing and re-balancing principles on the AMF \cite{3GPP2020}}
        \item \textcolor{red}{Provision of congestion awareness on the CP handling MM messages, i.e., NAS \cite{3GPP2020}}
        \item \textcolor{red}{Introduction of beam level MM support \cite{3GPP2020b}}
        \item \textcolor{red}{Intra-RAT (5GC to 5GC) and Inter-RAT (5GC to EPC and vice versa) HO support \cite{3GPP2020, 3GPP2020a}}
        \item \textcolor{red}{Well defined EPC and 5GC inter-working interface, i.e., N26 \cite{3GPP2020, 3GPP2020a}}
        \item \textcolor{red}{Mobility support at the D2D level \cite{Terrestrial2013, Jung2016}}
    \end{itemize}
    \item \textcolor{red}{3GPP 5G MM Cons}
    \begin{itemize}
        \item \textcolor{red}{Handover signaling in the CN is extremely sub-optimal \cite{Jain2019}}
        \item \textcolor{red}{RAT selection still relies on received signal quality fundamentals only \cite{3GPP2020b}}
        \item \textcolor{red}{A unified framework for cross-layer mechanisms, such as MPTCP-SCTP (transport layer), IPv6 multi-homing (network layer) and MR-DC (Physical and MAC layer) working together, has not been provisioned}
        \item \textcolor{red}{In IPv6 multi-homing, a single point of failure (SPoF) still exists, as the multiple PDU session anchors are still connected to a single UPF from where the paths branch out \cite{3GPP2020}}
        \item \textcolor{red}{Co-ordination between D2D peers for enacting an efficient MM strategy is not explored explicitly in the standards}
    \end{itemize}
\end{itemize}

\textcolor{red}{From the \textit{pros} and \textit{cons}, it can be deduced that the 3GPP 5G MM mechanisms will be able to support reliability parameters \textit{RL1} (owing to the support for MR-DC and IPv6 multi-homing, and hence, redundancy in the number of connections and flows), \textit{RL2} (owing to the support for MR-DC and handover procedures defined, thus ensuring seamless handover capability), \textit{RL3} (owing to managing mobility at the access, core and extreme edge network levels as well as local breakouts, thus introducing decentralization) and \textit{RL5} (owing to the congestion awareness feature in NAS). Next, for the flexibility parameters, 3GPP 5G MM mechanisms satisfy \textit{FL1} (owing to the granularity of service support per PDU session as well as per mobility level, and the ability to support on-demand MM through network slicing support), \textit{FL2} (owing to the ability to connect to multiple APs through MR-DC and IPv6 multi-homing support), \textit{FL3} (owing to the handover support at the access, core and extreme edge network levels via the Xn handover, N2 handover and 3GPP ProSe, respectively ) and \textit{FL5} (owing to the ability to take into account the context of the tenant via network slicing). Lastly, 3GPP 5G MM mechanisms, for the scalability criterion, satisfy parameters \textit{SL1} (owing to the AMF load balancing strategies, local breakout strategies, multi-level handover support as well as the granularity in service per mobility levels), \textit{SL4} (owing to local breakout and support for edge computing, thus leading to decentralization) and \textit{SL5} (since these are standards, implementation and integration is not a bottleneck).} 

\textcolor{red}{Note that, scalability parameters \textit{SL2} and \textit{SL3} are not supported owing to the sub-optimality in CN handover signaling as well as the presence of SPoFs, as stated in the \textit{cons} for the 3GPP 5G MM mechanisms. Also, given that the 3GPP 5G MM mechanisms provision both CP and DP related strategies as well as the core, access and extreme edge network related mechanisms, in Figure 2 they have been classified as illustrated.}

\subsection{\textcolor{red}{Other Research Efforts: Core, Access and Extreme Edge Network Solutions}}
From the perspective of MM strategies in 5G networks, the main objective of the ongoing academic and industrial research efforts has been to provision mechanisms that cater to the myriad user mobility and application profiles, as well as to ensure context/on-demand based service provision and continuity \cite{Kantor2015}. \textcolor{red}{For example, in \cite{Gramaglia2016}, a wide swathe of avenues that exist in the 5G MM design have been explored. It discusses an SDN based framework that can encompass strategies and techniques which grant certain level of adaptability (feedback based), flexibility (in terms of granularity provisions) and reliability (through availability of multiple paths) for 5G MM solutions.} Notably, and apart from the aforementioned broad study, specific areas of MM have also been tackled through research efforts such as \cite{Jain2019} wherein optimal handover signaling strategies for 4G-5G networks have been proposed.

\textcolor{red}{Hence, given that we will be analyzing a wide range of mechanisms and strategies, we have broadly classified them as being \emph{Core Network}, \emph{Access Network} and \emph{Extreme Edge Network} based solutions, as shown in Figure 2. These classifications reflect the regions in the network where the respective mechanisms generate the most impact. Concretely, \emph{Core network} based solutions will invoke solutions that primarily assist in the provision of MM services through the core network. Similarly, the \emph{Access network} and \emph{Extreme Edge network} solutions assist in provision of MM services through the access and extreme edge portion of the wireless network. And so, we now present a detailed discussion of these solutions alongside their efficacy in satisfying the criteria listed in Table 2.}

\subsubsection{Core Network Solutions}
\paragraph{\textcolor{red}{Discussion\\}} 
Core network solutions have been categorized further as either being \emph{SDN}, \emph{DMM} or \emph{Edge clouds} based. Solutions that utilize SDN to implement MM can be equipp\-ed with a global or locally-global network view. This top-view of the network enables MM solutions to offer a high degree of optimality. However, as a result of the convoluted 5G network scenario, the design of SDN CP also becomes increasingly crucial. Hence, the placement of SDN controller(s) (SDN-C) in the overall network topology is an important factor to consider \cite{GSchulz}. Consequently, we present a brief discussion on the SDN based solutions, which might be Centralized, Semi-Centralized or Hierarchical \cite{Li2014, Meneses2018, Assefa2017}. 

A centralized MM solution will consist of a single global SDN-C which monitors and manages the entire network. With the global view, it enables the formulation of optimal MM solutions. However, the centralized nature might not offer the scalability and reliability (SDN-C can be a SPoF) \cite{Li2014,Basloom} needed by 5G MM solutions. Note that, even though SDN-Cs might appear as SPoFs, corresponding clustering for load sharing and redundancy can help alleviate this issue. Specifically, and similar to the method proposed by 3GPP to pool the Mobility Management entities (MMEs) to avoid SPoF problem and to share the workload between MME instances, SDN-Cs can be clustered together to provision redundancy (and hence no SPoF) and workload sharing. Next, semi-centralized approaches divide the entire geographical region into smaller domains, each managed by a separate SDN-C. This SDN-C, responsible for handling MM in its domain, helps to enhance the network scalability. However, since each domain still has a single SDN-C managing it, SPoF issue might become a limiting factor. Further, for inter-domain HO, extensive signaling would be required between two SDN-Cs whilst the same would be non-existent in a centralized approach \cite{Li2014}. On account of this trade-off, a semi-centralized approach can be successful if an appropriate number of SDN domains are created, which do not increase the signaling burden while reinforcing the network reliability and scalability characteristics \cite{Basloom}. A combination of the aforementioned approaches, i.e., hierarchical approach, consists of SDN-Cs at multiple levels \cite{Li2014}. Whilst the global SDN-C behaves as a master (tuning HO parameters, manage inter domain HOs, etc.), the SDN-Cs in the lower hierarchical levels manage MM within their domains and function as slaves. Such an approach can hence provide the scalability and reliability required by 5G MM solutions.

Next, similar to the SDN based solutions, DMM based approaches will contribute significantly to the design and functioning of 5G networks. 
With the ability to provide a distributed DP in conjunction with a distributed/centralized CP \cite{Yang2016, Nguyen2016, Elgendi2016, Battulga2017, Liu2015}, DMM can enhance the scalability (by removing anchors prevalent in current MM solutions, i.e., decentralization) and flexibility (by allowing the most optimum access router for each flow independently) of the 5G networks. These approaches can be classified as being fully distributed, partially distributed and SDN based. 

The fully distributed approach whilst ensuring reliability and scalability by distributing both DP and CP, will encounter extensive amount of handover signaling between access routers (ARs) during a mobility event \cite{Yang2016, Nguyen2016}. \textcolor{red}{Note that the DP functionalities and location of ARs are the same as that of the UPFs. However, depending on the type of DMM approach, the CP is fully or partially located on the ARs themselves, instead of being located in a centralized controller.} And so, while the fully distributed approach is challenged by the signaling between ARs, the partially distributed (P-DMM) approach centralizes the CP, hence, alleviating this concern \cite{Yang2016, Nguyen2016}. The P-DMM approach also maintains the benefit of avoiding a single mobility anchor. However, an enhancement of this approach is the SDN based approach. Similar to the P-DMM approach, the CP is still at a central controller, i.e., SDN-C, however, the signaling between the controller and the DP devices is far more simplified as compared to the partially distributed approach. The reason being, in an SDN based approach, the ARs are converted to mere forwarding devices and it is the SDN-C that orchestrates the forwarding rules (routing table) on them to realize the data paths for the existing sessions in the network. Concretely, in the SDN based approach the DP devices no longer need to perform a handshake, like in the P-DMM approach, with the central controller to establish a route, instead the routing information is now fed to the DP devices by the SDN controller \cite{Nguyen2016, Elgendi2016}. These enhancements are further quantified in \cite{Nguyen2016} by the fact that the mean HO latency for SDN based DMM is reduced by 3.94\% as compared to P-DMM, while the E2E delay is reduced by 39.55\%. 

\textcolor{red}{Subsequent to these discussions, and given that the current standardization in 5G \cite{3GPP2020, 3GPP2020a} stipulates the functionality for mobility management to be split up between the AMF and the SMF NFs, it is noteworthy that the decoupling of the CP and DP and subsequent utilization of the aforesaid NFs via an SDN-C can provision the capability to implement fast and efficient MM solutions for 5G and beyond networks. Such solutions, on the basis of the discussions thus far, will be reliable, flexible and --to an extent-- scalable. Since, CN signaling during mobility events will still be a challenge, given the future network scenario, there remains a possibility for the SDN and DMM based 3GPP 5G MM solutions to be rendered sub-optimal.}

Lastly, edge clouds, which essentially refer to data clouds/processing centers close to the RAN within a given network infrastructure, can have a profound impact on the user QoS during mobility scenarios  (through fast access to data and compute resources) \cite{Li2014a}. Henceforth, several studies such as \cite{Urgaonkar2015,Wang2018,Architecture2016,Machen2018,Mtibaa2018, Mach2017} alongside 3GPP and ETSI \cite{ETSI2018}, have studied the fundamental concepts of utilizing the edge clouds for fast data access (via data caching) as well as for processing capabilities (i.e., performing certain MM operations without the messages having to traverse the entire CN). Note that, we classify the edge clouds to be a CN solution, even though we state that they are most likely to be closer to the RAN, because, certain topology designs might entail a hierarchical setup. In this hierarchical setup, there will be some edge clouds that are placed close to the RAN and some of them being placed further away from the RAN, say close to the S-GW and Packet Gateway (P-GW) in an LTE network \cite{Li2014a}. Such an approach can help in caching data according to their level of popularity, taking into account CN traffic as well as the latency to retrieve the requested content \cite{Li2014a}. 


\paragraph{\textcolor{red}{Analysis}\\}
\textcolor{red}{For analyzing the core network solutions we utilize the generic classifications, i.e., SDN based, DMM based and Edge Cloud solutions, and firstly list their \textit{pros} and \textit{cons}.} 

\begin{itemize}
    \item \textcolor{red}{SDN based mechanism Pros}
    \begin{itemize}
        \item \textcolor{red}{Provisions global view of the network \cite{Li2014, Assefa2017}}
        \item \textcolor{red}{Provisions hierarchical solutions, thus enabling decentralization \cite{Li2014}}
        \item \textcolor{red}{Provisions the ability to manage CN signaling, and hence, DP paths during mobility events \cite{Li2014, Meneses2018, Assefa2017}} 
        \item \textcolor{red}{Provisions a single point of collection for network statistics thus enabling the design and development of context based MM mechanisms \cite{Basloom}}
    \end{itemize}
    \item \textcolor{red}{SDN based mechanism Cons}
    \begin{itemize}
        \item \textcolor{red}{Extensive CN signaling for managing handovers in a centralized/semi-centralized approach \cite{Li2014}}
        \item \textcolor{red}{It does not alleviate the issue of mobility anchors which can lead to SPoFs in the DP}
    \end{itemize}
    \item \textcolor{red}{DMM based mechanism Pros}
    \begin{itemize}
        \item \textcolor{red}{Provision decentralization of the mobility management anchors \cite{Yang2016, Nguyen2016, Elgendi2016, Battulga2017}}
        \item \textcolor{red}{Assist the CN in implementing efficient data paths for UEs undergoing mobility \cite{Nguyen2016, Liu2015, Yang2016}}
    \end{itemize}
    \item \textcolor{red}{DMM based mechanism Cons}
    \begin{itemize}
        \item \textcolor{red}{Fully decentralized solution introduces extensive CN signaling in order to manage the changes in data paths and mobility anchors, and hence, handovers \cite{Nguyen2016}}
        \item \textcolor{red}{Partially distributed solution, while solving the extensive CN signaling, introduces a central controller, and hence, an SPoF \cite{Nguyen2016}}
        \item \textcolor{red}{Co-existence and integration with already deployed networks and devices will be a significant challenge \cite{Liu2015}}
    \end{itemize}
    \item \textcolor{red}{Edge clouds Pros}
    \begin{itemize}
        \item \textcolor{red}{Ensure data offloading opportunities, and hence, reduction in CN traffic load \cite{ETSI2018, Urgaonkar2015}}
        \item \textcolor{red}{Facilitate processing of MM related tasks without the messages having to traverse the CN \cite{Mtibaa2018}}
        \item \textcolor{red}{Context awareness \cite{Mtibaa2018,Mach2017}}
    \end{itemize}
    \item \textcolor{red}{Edge clouds Cons}
    \begin{itemize}
        \item \textcolor{red}{Require dedicated infrastructure and appropriate placement \cite{ETSI2018, Urgaonkar2015, Machen2018}}
        \item \textcolor{red}{Require fast service migration strategies to ensure seamless mobility \cite{Wang2018}}
    \end{itemize}
\end{itemize}

\textcolor{red}{From these \textit{pros} and \textit{cons} as well as the preceding discussions, it is evident that the SDN based solutions satisfies parameter \textit{RL2} (allowing for seamless mobility), \textit{RL3} (through the provision of decentralized solutions), \textit{RL4} (through the ability to re-program paths in CN via orchestration of OF rules) and \textit{RL5} (through the ability to utilize network statistics for traffic steering with the CN) for the reliability criterion. For the flexibility criteria, the SDN based mechanisms satisfy the parameters \textit{FL1} (through the capability of orchestrating policies dependent on flow type, slice, etc.), \textit{FL3} (by allowing for CN based MM solutions that will work in synergy with the access network based solutions) and \textit{FL4} (through the global view of the network wherein a variety of parameters such as network load, QoS requirements, etc., are considered). In terms of scalability, SDN based solutions satisfy parameters \textit{SL1} to \textit{SL3} (given the ability to manage and steer traffic flows with the ability of having a distributed, hierarchical or centralized implementation) and \textit{SL4} (due to the possibility of having a decentralized configuration).} 

\textcolor{red}{The DMM based solutions, however only satisfy parameters \textit{RL2} (allowing for seamless handovers) and \textit{RL3} (due to the decentralized nature) in the reliability criterion. Further, for the flexibility criterion, DMM based solutions only satisfy parameter \textit{FL1}, i.e., they only offer granularity of service by preventing any mobility anchor. It is noteworthy though that, from the scalability aspect DMM based solutions, like SDN based solutions, satisfy parameters \textit{SL1} to \textit{SL4}, and for the same reasons.} 

\textcolor{red}{Lastly, for the edge cloud based solutions, parameters \textit{RL2} (allowing for seamless mobility through fast access to data/processing capabilities upon migration to the target network) and \textit{RL3} (allowing decentralization of MM based services) are satisfied for the reliability criterion. For the flexibility criteria, parameters \textit{FL1} (due to the ability to provision services based on mobility and application profiles), \textit{FL3} (by allowing for MM methods at the edge network level in addition to the access and core network based solutions), \textit{FL4} (by provisioning processing capabilities for user association/AP selection services) and \textit{FL5} (by allowing for context awareness in data caching according to user mobility) are satisfied. Additionally, for the scalability criteria, parameters \textit{SL1} to \textit{SL4} are satisfied by the edge cloud solutions. The reason being, they allow for decentralization which can consequently permit better capability to manage connections and control messages due to increasing number of users.} 

\textcolor{red}{It is important to state here that, given the SDN based mechanisms assist in MM through CP procedures, DMM based solutions assist through CP procedures as well as provision alternate and effective DP paths, and Edge clouds provision alternate and effective data paths, they have been classified as being CP, CP/DP and DP procedures, respectively, in Figure 2.}

\subsubsection{Access Network Solutions}

\paragraph{\textcolor{red}{Discussion}\\} 
As part of access network strategies, one of the key approaches that has been proposed, and similar to LTE dual connectivity, is the concept of phantom cell \cite{Nakamura2013}. It allows the UE to camp its CP on a MC, while its DP is being handled at the small cells that lie within the coverage of the earlier mentioned MC. This, in essence offers a low signaling cost regime to perform the intra-MC HOs as the UE does not need to access the CN for radio resource management operations during HO. Concretely, the MC handles the radio resource allocation operations for the phantom cells, and hence, during HOs between the phantom cells the CN signaling is avoided \cite{Carrier2015}.

Moreover, owing to the softwarization of the complete network, the process of exchanging information between the various OSI layers, i.e., implementation of the cross layer strategy, is eased. This in turn allows the network to formulate solutions that are optimal, taking into cognizance the impact and benefits that the solution will produce at various levels of the network \cite{Emam2020, Emam2020a, Al-rubaye2016}. \textcolor{red}{However, to realize cross-layer techniques, significant modifications to the software architecture of the protocol stack will be necessary \cite{Emam2020, Emam2020a, Al-rubaye2016}.} Another consequence of the softwarization process is the RAN-as-a-Service (RANaaS), also known as Cloud-RAN (C-RAN), which allows on-demand allocation of access network resources (e.g., Baseband unit (BBU) pool, BBU- Remote Radiohead (RRH) functional splitting) depending on the network and user context \cite{Nikaein2015, Outtagarts2015, Sabella2013}. Additionally, the BBU pool, through close interaction of various RATs at a single location, can orchestrate fast handovers on-demand \cite{Liu2012}. 

\indent However, in order to choose the best APs to connect to in a multi-RAT scenario, computationally tractable RAT selection mechanisms need to be adopted. The multi-RAT solutions are a broad classification for the myriad RAT selection processes (Optimization based, Fuzzy logic and Genetic Algorithm based, RSSI based, etc. \cite{Zekri2012, Passast2019, Goudarzi2019, Wang2019}) that have been proposed. From our earlier discussions it is evident that RSSI based methods, although simple, do not weigh in other parameters such as network load, backhaul conditions, or user/network policies, for a RAT selection decision. This will most certainly result in sub-optimal solutions. But, optimized mechanisms, that can facilitate closed form solutions and are computationally tractable, will be able to capture more features from the network. Consequently, context aware mechanisms, such as \cite{Calabuig2017,jain2020user}, will lead to optimal solutions that can be implemented for real-time scenarios. 

It must be stated here that, the aforesaid HO decision may be executed either at the UE (user-centric) \cite{Calabuig2017}, at the network, or as a joint effort between the UE and the network (hybrid decision process).

\paragraph{\textcolor{red}{Analysis}\\} 

\textcolor{red}{As part of the analysis for the access network solutions, we firstly present the \textit{pros} and \textit{cons} for each mechanism discussed above, as follows:}

\begin{itemize}
    \item \textcolor{red}{Phantom Cell method Pros}
    \begin{itemize}
        \item \textcolor{red}{Grants the ability to a UE to connect to multiple APs simultaneously, thus also granting redundancy in physical layer connections \cite{Nakamura2013}}
        \item \textcolor{red}{Provisions the ability to allow per-flow and per-user granularity of service \cite{Nakamura2013, Jain2017}}
        \item \textcolor{red}{Handover support at access network level \cite{Nakamura2013}}
        \item \textcolor{red}{Ease of implementation due to existing standards on MR-DC \cite{3GPP2020,Nakamura2013}}
    \end{itemize}
    \item \textcolor{red}{Phantom Cell method Cons}
    \begin{itemize}
        \item \textcolor{red}{Handovers between different MC domains will still entail service disruption \cite{3GPP2020,Nakamura2013}}
        \item \textcolor{red}{Inter-MC domain handover signaling will still be a significant burden on the CN \cite{Nakamura2013,Jain2019}}
    \end{itemize}
    \item \textcolor{red}{RANaaS Pros}
    \begin{itemize}
        \item \textcolor{red}{Provisions on-demand allocation of network resources at the RAN level \cite{Sabella2013, Nikaein2015, Outtagarts2015}}
        \item \textcolor{red}{Provisions the ability to execute on-demand handovers, through close interaction between the various RATs that are integrated at a BBU pool \cite{Liu2012}}
        \item \textcolor{red}{Assists in allowing UEs to camp on more than one AP}
        \item \textcolor{red}{Introduces support for executing handovers at the access network level \cite{Liu2012}}
        \item \textcolor{red}{Introduces the ability to utilize per-flow/channel granularity of service by being able to manage the physical connections more centrally \cite{Liu2012,Nikaein2015,Sabella2013,Outtagarts2015}}
    \end{itemize}
    \item \textcolor{red}{RANaaS Cons}
    \begin{itemize}
        \item \textcolor{red}{Requires a complete architectural overhaul at the RAN side of the network \cite{Sabella2013, Nikaein2015, Outtagarts2015}}
    \end{itemize}
    \item \textcolor{red}{Cross layer Pros}
    \begin{itemize}
        \item \textcolor{red}{Allows for the sharing of network statistics between the various OSI layers \cite{Emam2020, Emam2020a, Al-rubaye2016}}
        \item \textcolor{red}{Allowing for interaction between multiple OSI layers, thus facilitating the possibility of efficient utilization of multi-homing \cite{Emam2020, Emam2020a, Al-rubaye2016, ITU-T2011}}
    \end{itemize}
    \item \textcolor{red}{Cross layer Cons}
    \begin{itemize}
        \item \textcolor{red}{Requires significant software modifications to the existing modular nature of the protocol structure \cite{Emam2020, Emam2020a, Al-rubaye2016}}
    \end{itemize}
    \item \textcolor{red}{Intelligent RAT selection Pros}
    \begin{itemize}
        \item \textcolor{red}{Optimized RAT Selection strategies \cite{Zekri2012, Passast2019, Goudarzi2019, Wang2019, Calabuig2017}}
        \item \textcolor{red}{Utilization of parameters such as AP load, UE context, etc., for RAT selection \cite{Zekri2012, Passast2019, Goudarzi2019, Wang2019, Calabuig2017}}
        \item \textcolor{red}{Provisioning the ability to select RATs per-slice/user/flow \cite{Calabuig2017}}
        \item \textcolor{red}{Provisioning the ability to select multiple APs (possibly belonging to multiple RATs) \cite{jain2020user}}
    \end{itemize}
    \item \textcolor{red}{Intelligent RAT selection Cons}
    \begin{itemize}
        \item \textcolor{red}{Requires rapid collection of network statistics to perform well informed selection}
        \item \textcolor{red}{Computational complexity and convergence time of RAT selection algorithms will be critical, given the QoS requirements in 5G \cite{jain2020user}}
    \end{itemize}
\end{itemize}

\textcolor{red}{Given the discussions in Section 5.2.2.1  and the \textit{pros} and \textit{cons} listed above, we now determine the parameters, listed in Table 2, satisfied by each of the mechanisms explored. Concretely, for the phantom cell method, parameters \textit{RL1} (redundancy in physical layer connections) and \textit{RL2} (seamless mobility) are satisfied for the reliability criterion. For the flexibility criterion, parameters \textit{FL1} (by permitting the possibility of per-flow and per-user based MM), \textit{FL2} (allowing for connectivity to multiple APs potentially belonging to different RATs) and \textit{FL3} (provisioning handover support at the access network level that will work in synergy with CN based mechanism) are satisfied. In terms of scalability, the phantom cell method satisfies parameters \textit{SL1} to \textit{SL3} (owing to the handling of handover related computation and decision at the access network) and \textit{SL5} (owing to the existing standards on MR-DC, as discussed in Section 3).} 

\textcolor{red}{Next, the RAN-as-a-service concept satisfies parameters \textit{RL2} (allowing for seamless  handovers) and \textit{RL5} (the softwarized nature enables dynamic initiation for RAN functionality such as BBU resources, functional splits, etc., depending on the network and user context) for reliability, parameters \textit{FL1} (allowing for per-flow, per-user, per-slice, etc., service granularity through its softwarized nature), \textit{FL2} (allowing the possibility for connecting a user to multiple APs through its softwarized nature), \textit{FL3} (provisioning handover support at the access network which will work in synergy with the CN and edge network based methods) and \textit{FL4} (enabling the possibility of collection and utilization of RAN based information and generating intelligent AP selection/user association decisions) for flexibility, and parameters \textit{SL1} to \textit{SL3} (by offloading handover decision making and signaling to the access network) for scalability.} 

\textcolor{red}{On the other hand, the cross-layer method only satisfies parameters \textit{RL2} (allowing for seamless handover) and \textit{RL5} (allows for congestion aware method by sharing statistics about queue lengths, buffer sizes, etc., amongst the various layers) for the reliability criteria. Further, for the flexibility criteria it satisfies only parameters \textit{FL2} (by allowing for the possibility of multi-homing, etc.) and \textit{FL4} (allowing for the possibility of sharing statistics and other information amongst the various OSI layers and enabling joint optimization for AP selection, path re-routing, etc.).} 

\textcolor{red}{Lastly, for the intelligent RAT selection methods parameter \textit{RL2} (allowing for seamless handover through optimized decisions on RAT selection) is satisfied for the reliability criterion. For the flexibility criterion, parameters \textit{FL1} (allowing for the possibility of flow/user/slice based RAT selection), \textit{FL2} (allowing for the possibility to select multiple RATs for a given user) and \textit{FL5} (via the ability to utilize user and network context for RAT selection) are satisfied, while for scalability only parameter \textit{SL5} (owing to the extensive body of research for optimal RAT selection strategies) is satisfied.} 

\textcolor{red}{It is important to state here that, given the intelligent RAT selection mechanism assists in MM through RAT selection (which is a CP task) and provision of effective and alternate DP paths, the phantom cell method provisions support for MM by handling the CP signaling for SC selection as well as provision alternate and effective DP paths via SCs, and RANaaS and Cross layer strategies assist through efficient resource allocation decisions (which is a CP task), thus they have been classified as being CP/DP, CP/DP, CP and CP procedures, respectively, in Figure 2.}

\subsubsection{Extreme Edge Network Solutions}
\paragraph{\textcolor{red}{Discussion}\\}
Contrasting to the design and implementation of access and core network based methods, the extreme edge network based solutions consider the potential of utilizing D2D techniques for facilitating seamless HO. \textcolor{red}{Multiple research efforts, such as \cite{Yilmaz2014a,Ouali2020,Klempous2020,Barua2017, Barua2016}, have provisioned methodologies to handle mobility of D2D pairs. Concretely, in \cite{Yilmaz2014a} two types of handovers for D2D pairs have been provisioned. These are either \textit{D2D aware} and \textit{D2D triggered} handovers. They take into account the fact that the control of the D2D pair can be handed over independently of the actual cellular handover. And so, for the \textit{D2D aware} handover, the D2D pair control (and if possible the cellular control) is handed over from the source eNB to the target eNB only after both the devices in the D2D pair satisfy the conditions to handover to the target eNB. On the other hand, the \textit{D2D triggered} handover mechanism aims at clustering the devices of a D2D group in minimum number of cells. Hence, during mobility events the algorithm tries to determine the cell to which the majority of devices within the D2D group belong too.}

\textcolor{red}{Similarly, in \cite{Ouali2020} two handover management mechanisms have been proposed. While the joint handover strategy aims at migrating both the devices in a D2D pair simultaneously to the target eNB, the half handover stipulates that such a migration can be asynchronous. Furthermore, the D2D handover decision has also been specified in \cite{Ouali2020}. The Channel Quality Information (CQI) criteria has been utilized for the same. Next, in \cite{Klempous2020}, a markov chain based model has been proposed for D2D mobility.}

\textcolor{red}{Lastly, the work done in references \cite{Barua2017, Barua2016} develops a model and simulation framework analyzing D2D mobility. Specifically, it considers a D2D pair with one of them being a transmitter (TX) and the other being just a receiver (RX). Thus, a handover procedure is defined for the scenario when the TX moves to the target eNB. In this procedure, the control of the D2D pair is transferred to the target eNB as soon as the TX migrates to it.}


\paragraph{\textcolor{red}{Analysis}\\}
\textcolor{red}{We firstly present the \textit{pros} and \textit{cons} for the D2D strategies as follows:}

\begin{itemize}
    \item \textcolor{red}{D2D strategy Pros}
    \begin{itemize}
        \item \textcolor{red}{Provisions D2D handover management strategies \cite{Yilmaz2014a,Ouali2020, Barua2016}}
        \item \textcolor{red}{Provisions MM support at the extreme edge network level \cite{Yilmaz2014a,Ouali2020,Klempous2020,Barua2017, Barua2016}}
        \item \textcolor{red}{Provisions the ability to decentralize MM functionality\\ \newline}
    \end{itemize}
    \item \textcolor{red}{D2D Strategy Cons}
    \begin{itemize}
        \item \textcolor{red}{Control signaling overhead will be a challenge \cite{Ouali2020, Yilmaz2014a}}
        \item \textcolor{red}{The viability with regards to energy efficiency of D2D peers as well as latency incurred in conveying the decisions with regards to MM are un-explored questions}
    \end{itemize}
\end{itemize}

\textcolor{red}{Based on the discussions and the aforesaid \textit{pros} and \textit{cons}, the device-to-device methods satisfy parameter \textit{RL2} (through the provision of various seamless handover management studies) for reliability, parameter \textit{FL3} (provisioning mobility support at the edge network level which will work in synergy with access and core network based methods) for flexibility, and parameter \textit{SL4} (allowing for the decentralization of MM functionality) for scalability.}


\textcolor{red}{Note that, given the D2D mechanism assists in MM through provision of CP assistance, thus they have been classified as being CP procedure in Figure 2.}

\subsection{B5G Networks}

\textcolor{red}{In addition to the discussions in Sections 5.1 and 5.2, in this section we present a short study detailing the challenges that current state-of-the-art mechanisms will continue to face for B5G networks. Furthermore, given the special characteristics that B5G networks will pose, as shown in Figure 1, we also list potential research areas for MM in B5G networks. Note that, these are then utilized in the subsequent section wherein we define challenges and potential solutions for 5G and beyond MM.}

\textcolor{red}{Concretely, while \emph{SDN} and \emph{NFV} will provide the tools for the B5G networks to provision rapid programmability of the meta-surfaces, during mobility scenarios they will be challenged critically. The reason being that, while current networking paradigms permit anywhere between 1 ms--10 ms time interval for performing any programmability task (latency restrictions, as specified in current 5G networks \cite{Parvez2018}, on most services), in B5G networks this will be constrained even further as additional surfaces need to be programmed and orchestrated. Specifically, an increased number of surfaces/network nodes leads to more data required to be processed for generating appropriate programmability decisions. These decisions then need to be sent out (orchestrated) to the relatively large number of network nodes (including meta-surfaces), to execute the given task. Hence, this leads to an increased latency constraint on the network programmability aspect. Further, while the meta-surfaces provide a higher degree of freedom to the operator, they need to be programmed, as mentioned above. This introduces the challenging aspect of managing the SDN domains, NFV orchestration and the related signaling. As a consequence, the compactness as well as the efficiency of the current state-of-the-art SDN and NFV procedures will be challenged.}

\textcolor{red}{Next, with techniques such as DC, the challenge will be multi-fold as B5G networks will not just comprise of meta-surfaces, which can also act as a MIMO array, but they will also be equipped with Terahertz and mobile AP based multi-tier networks. And while, DC and multi-RAT procedures, as stated in Sections 5.1 and 5.2, will aid in ensuring a context-aware network selection procedure, the complexity for the access network techniques will be compounded by the fact that not only will they need to ensure QoS requirements, but they will have to also ensure sufficient available access bandwidth as well as backhaul bandwidth. Note that with the backhaul bandwidth there will be a significant design challenge since VLC technology is capable of carrying data rates of up to 1 Tbps. Current backhaul technologies cannot provision such high bandwidths \cite{Jaber2016b}. Further, it is important to reiterate that the network will be composed of not only 4G-LTE and mmWave APs, but there will also be VLC and drone based APs, which essentially are the main reason for the increased complexity as discussed above.}

\textcolor{red}{Moreover, for the edge clouds, while they aid in allowing low latency access to cached content as well as the compute resources, the deployment strategies will need to be rethought given the ongoing growth pattern for data usage as well as the number of served devices coupled with more resource hungry services. Certain important recent studies in this direction have been provisioned via references \cite{Santoyo-Gonzalez2018, Leyva-Pupo2019}.}

\textcolor{red}{Given these significant shortcomings in the current state-of-the-art mechanisms towards B5G networks as well as taking into account the seminal works in the area of B5G techniques \cite{Boulogeorgos2018, Renzo2019, Basar2019, Chowdhury2018} \cite{Sekander2018}, the potential areas of research in MM for these networks are as follows:}

\begin{itemize}
    \item Characterization of the channel between meta-surface and the users, and meta-surface and the AP, in the event of user/AP being mobile, for the purpose of MM decisions
    \item Consideration of reliability and coverage of VLC link for MM decisions
    \item Characterization of the computational complexity for re-calibrating the meta-surfaces alongside the network, during mobility events
    \item Impact of mobility upon the programmable environment\footnotemark[1] \footnotetext[1]{By environment, we refer to the physical environment that lies between the transmitter and receiver.} concept, drone based communication and VLC
    \item Optimal RAT and AP selection with a programmable environment
    \item Optimal RAT and AP selection in scenarios where both the UE and AP (drone based) are mobile
    \item Characterizing the computational complexity of optimization methodologies for user association
    \item Methods to handle possible increase in handover signaling/messaging during other network processes, such as reprogramming meta-surfaces to serve mobile users
    \item Formulation of a sound heterogenous RAT strategy, just like the 4G-5G concept, given mmWave and Terahertz technologies and their associated challenges related to coverage.
\end{itemize}

Note that, the aforementioned research areas do not form an exhaustive list, but are broadly indicative of what aspects remain to be explored with regards to MM in B5G networks. \\ \newline

\begin{sidewaystable*}
\renewcommand{\arraystretch}{1.1}
\caption{\textcolor{red}{Compliance with the Reliability, Scalability and Flexibility criteria of the Current state-of-the-art MM mechanism/standard}}
\centering
\color{red}\begin{tabular}{|*{10}{>{\centering\arraybackslash}m{0.6 cm}|>{\centering\arraybackslash}m{1.1 cm}|}}
\cline{3-20}
\multicolumn{2}{c|}{} & \multicolumn{2}{c|}{} & \multicolumn{16}{c|}{\textbf{Other Research Efforts}}\\ \cline{5-20}
\multicolumn{2}{c|}{} & \multicolumn{2}{m{2cm}|}{\textbf{3GPP 5G MM mechanism}} & \multicolumn{6}{c|}{\textbf{Core Network Solutions}} & \multicolumn{8}{c|}{\textbf{Access Network Solutions}} & \multicolumn{2}{m{2cm}|}{\textbf{Extreme Edge Network Solutions}} \\ \cline{5-20}
\multicolumn{2}{c|}{} & \multicolumn{2}{c|}{} & \multicolumn{2}{m{1.5cm}|}{\textbf{SDN based}} & \multicolumn{2}{m{1.5cm}|}{\textbf{DMM based}} & \multicolumn{2}{m{1.5cm}|}{\textbf{Edge Clouds}} & \multicolumn{2}{m{1.5cm}|}{\textbf{Phantom Cell Method}} &  \multicolumn{2}{m{1.5cm}|}{\textbf{RAN-as-a-Service}} & \multicolumn{2}{m{1.5cm}|}{\textbf{Cross layer}} & \multicolumn{2}{m{1.5cm}|}{\textbf{Intelligent RAT selection}} & \multicolumn{2}{m{1.5cm}|}{\textbf{Device-to-Device}} \\ \hline
\multicolumn{2}{|c|}{} & Cnf.$^\dagger$ & Refs.$^\delta$ & Cnf. & Refs. & Cnf. & Refs. & Cnf. & Refs. & Cnf. & Refs. & Cnf. & Refs. & Cnf. & Refs. & Cnf. & Refs. & Cnf. & Refs. \\ \hline

\multirow{5}{*}[-1em]{\rotatebox{90}{\textbf{Reliability}}} & \textbf{RL1} & \Checkmark & \multirow{5}{*}[0.1em]{\cite{3GPP2020,3GPP2020a}} &$\times$ & \multirow{5}{*}[0.1em]{\cite{Li2014,Meneses2018, Assefa2017, Basloom}} & $\times$ & \multirow{5}{*}[0.1em]{\cite{Liu2015,Yang2016}}  & $\times$ & \multirow{5}{*}[0.1em]{\cite{Urgaonkar2015}} & \Checkmark & \multirow{5}{*}[0.1em]{\cite{Nakamura2013}} & $\times$ & \multirow{5}{*}[0.1em]{\cite{Nikaein2015}} & $\times$ & \multirow{5}{*}[0.1em]{\cite{ITU-T2011,Emam2020}} & $\times$ & \multirow{5}{*}[0.1em]{\cite{Zekri2012, Passast2019}} & $\times$ & \multirow{5}{*}[1.5em]{\cite{Yilmaz2014a}} \\ \cline{2-3}\cline{5-5}\cline{7-7}\cline{9-9}\cline{11-11}\cline{13-13}\cline{15-15}\cline{17-17}\cline{19-19}
& \textbf{RL2} & \Checkmark & & \Checkmark & & \Checkmark & & \Checkmark & & \Checkmark & & \Checkmark & & \Checkmark& & \Checkmark & & \Checkmark & \\ \cline{2-3}\cline{5-5}\cline{7-7}\cline{9-9}\cline{11-11}\cline{13-13}\cline{15-15}\cline{17-17}\cline{19-19}
& \textbf{RL3} & \Checkmark & & \Checkmark & & \Checkmark & & \Checkmark & & $\times$ & & $\times$ & & $\times$ & & $\times$ & & $\times$ & \cite{Ouali2020} \\ \cline{2-3}\cline{5-5}\cline{7-7}\cline{9-9}\cline{11-11}\cline{13-13}\cline{15-15}\cline{17-17}\cline{19-19}
& \textbf{RL4} & $\times$ & \cite{Terrestrial2013,Jung2016} & \Checkmark & & $\times$ & \cite{Nguyen2016,Elgendi2016, Battulga2017} & $\times$ & \cite{Mtibaa2018,Mach2017,ETSI2018} & $\times$ & & $\times$ & \cite{Outtagarts2015,Sabella2013} & $\times$ & \cite{Emam2020a,Al-rubaye2016} & $\times$ & \cite{Goudarzi2019,Wang2019, Calabuig2017}& $\times$ & \cite{Barua2016, Barua2017} \\ \cline{2-3}\cline{5-5}\cline{7-7}\cline{9-9}\cline{11-11}\cline{13-13}\cline{15-15}\cline{17-17}\cline{19-19}
& \textbf{RL5} & \Checkmark & & \Checkmark & & $\times$ & & $\times$ & & $\times$ & & \Checkmark & & \Checkmark & & $\times$ & & $\times$ & \\ \hline \hline
\multirow{5}{*}[-1em]{\rotatebox{90}{\textbf{Flexibility}}} & \textbf{FL1} & \Checkmark & \multirow{5}{*}[1.5em]{\cite{3GPP2020, Terrestrial2013}} & \Checkmark & \multirow{5}{*}[0.1em]{\cite{Li2014,Meneses2018}}& \Checkmark & \multirow{5}{*}[0.1em]{\cite{Liu2015,Nguyen2016}} & \Checkmark & \multirow{5}{*}[0.1em]{\cite{Urgaonkar2015}} & \Checkmark & \multirow{5}{*}[0.1em]{\cite{Nakamura2013,Jain2017}}& \Checkmark & \multirow{5}{*}[0.1em]{\cite{Nikaein2015}} & $\times$ & \multirow{5}{*}[0.1em]{\cite{ITU-T2011,Emam2020}} & \Checkmark & \multirow{5}{*}[0.1em]{\cite{Passast2019}} & $\times$ & \multirow{5}{*}[1.5em]{\cite{Yilmaz2014a}} \\ \cline{2-3}\cline{5-5}\cline{7-7}\cline{9-9}\cline{11-11}\cline{13-13}\cline{15-15}\cline{17-17}\cline{19-19}
& \textbf{FL2} & \Checkmark & & $\times$ & & $\times$ & & $\times$ & & \Checkmark & & \Checkmark & & \Checkmark & & \Checkmark & & $\times$ &  \\ \cline{2-3}\cline{5-5}\cline{7-7}\cline{9-9}\cline{11-11}\cline{13-13}\cline{15-15}\cline{17-17}\cline{19-19}
& \textbf{FL3} & \Checkmark & \cite{3GPP2020a, 3GPP2020b} & \Checkmark & & $\times$ & & \Checkmark & & \Checkmark & & \Checkmark & & $\times$ & & $\times$ & & \Checkmark & \cite{Ouali2020, Klempous2020} \\ \cline{2-3}\cline{5-5}\cline{7-7}\cline{9-9}\cline{11-11}\cline{13-13}\cline{15-15}\cline{17-17}\cline{19-19}
& \textbf{FL4} & $\times$ & \cite{Jung2016} & \Checkmark & & $\times$ & & \Checkmark & \cite{Mtibaa2018,Mach2017,ETSI2018} & $\times$ & & \Checkmark & \cite{Outtagarts2015,Sabella2013,Liu2012} & \Checkmark & \cite{Emam2020a,Al-rubaye2016} & $\times$ & \cite{Goudarzi2019,Wang2019,Calabuig2017}& $\times$ & \cite{Barua2017, Barua2016} \\ \cline{2-3}\cline{5-5}\cline{7-7}\cline{9-9}\cline{11-11}\cline{13-13}\cline{15-15}\cline{17-17}\cline{19-19}
& \textbf{FL5} & \Checkmark& & $\times$ & & $\times$ & & \Checkmark & & $\times$ & & $\times$ & & $\times$ & & \Checkmark & & $\times$ & \\ \hline \hline
\multirow{5}{*}[-1em]{\rotatebox{90}{\textbf{Scalability}}} & \textbf{SL1} & \Checkmark & \multirow{5}{*}[0.1em]{\cite{3GPP2020, Terrestrial2013}} & \Checkmark & \multirow{5}{*}[0.1em]{\cite{Assefa2017, Basloom}}& \Checkmark & \multirow{5}{*}[0.1em]{\cite{Liu2015}}& \Checkmark & \multirow{5}{*}[0.1em]{\cite{Urgaonkar2015}}  & \Checkmark &  \multirow{5}{*}[0.1em]{\cite{Nakamura2013,3GPP2020}} & \Checkmark & \multirow{5}{*}[0.1em]{\cite{Nikaein2015}} & $\times$ & \multirow{5}{*}[0.1em]{\cite{Emam2020}} & $\times$ & \multirow{5}{*}[0.1em]{\cite{Zekri2012, Passast2019}}& $\times$ & \multirow{5}{*}[1.5em]{\cite{Yilmaz2014a}} \\ \cline{2-3}\cline{5-5}\cline{7-7}\cline{9-9}\cline{11-11}\cline{13-13}\cline{15-15}\cline{17-17}\cline{19-19}
& \textbf{SL2} & $\times$ & & \Checkmark & & \Checkmark & & \Checkmark & & \Checkmark & & \Checkmark & & $\times$ & & $\times$ & & $\times$ & \\ \cline{2-3}\cline{5-5}\cline{7-7}\cline{9-9}\cline{11-11}\cline{13-13}\cline{15-15}\cline{17-17}\cline{19-19}
& \textbf{SL3} & $\times$ & & \Checkmark & & \Checkmark & & \Checkmark & & \Checkmark & & \Checkmark & & $\times$ & & $\times$ & & $\times$ & \cite{Ouali2020,Klempous2020}\\ \cline{2-3}\cline{5-5}\cline{7-7}\cline{9-9}\cline{11-11}\cline{13-13}\cline{15-15}\cline{17-17}\cline{19-19}
& \textbf{SL4} & \Checkmark & \cite{3GPP2020b} & \Checkmark & & \Checkmark & \cite{Nguyen2016,Elgendi2016, Battulga2017}& \Checkmark & \cite{Mtibaa2018,Mach2017,ETSI2018} & $\times$ & & $\times$ & \cite{Outtagarts2015,Sabella2013,Liu2012} & $\times$ & \cite{Al-rubaye2016,Emam2020a} & $\times$ & \cite{Goudarzi2019,Wang2019,Calabuig2017}& \Checkmark & \cite{Barua2017,Barua2016}\\ \cline{2-3}\cline{5-5}\cline{7-7}\cline{9-9}\cline{11-11}\cline{13-13}\cline{15-15}\cline{17-17}\cline{19-19}
& \textbf{SL5} & \Checkmark & & $\times$ & & $\times$ & & $\times$ & & \Checkmark & & $\times$ & & $\times$ & & \Checkmark & & $\times$ & \\ \hline 
\multicolumn{20}{l}{$^{\dagger}$The conformance (Cnf.) of a given mechanism for a given criterion.} \\
\multicolumn{20}{l}{$^{\delta}$The corroborating references (Refs.), if any, for the specified conformance of a mechanism for a given criterion} 
\end{tabular}
\end{sidewaystable*} 


\noindent \textcolor{red}{To summarize, in this section we firstly introduced the 5G service based architecture and the classification of the various mechanisms that we analyzed, through Figure 2. Following this, we qualitatively analyzed the 3GPP 5G MM mechanisms as well as other research efforts with regards to their efficacy towards 5G and beyond MM solutions. Consequently, we introduce Table 4 wherein we indicate the parameters  that  each  of  the  explored  methods  satisfies  for  the  reliability,  scalability  and  flexibility  criteria (Table 2). We also enlist the important references that have lead us to the development of Table 4,  as presented in this article. And so, from the capability profiles of each mechanism, as illustrated in Table 4, it is evident that even after significant efforts none of them completely meet the specified requirements as expected for the 5G and beyond MM mechanisms. Concretely, neither the 3GPP 5G MM mechanisms nor the other academic and industrial research efforts satisfy all the criteria completely. Subsequently, it is deduced that none of the analyzed mechanisms satisfy the requirements for the future MM mechanisms, as listed in Table 1. Hence, through the aforesaid qualitative analysis we have further exposed the gaps in the design and development for 5G and beyond MM mechanisms.}

\section{Challenges, Potential Solutions and Future framework}


\textcolor{red}{From our discussions in Sections 2 to 5, we have highlighted the requirements from MM mechanisms as well as the criteria that future MM mechanisms should satisfy to meet these requirements in Tables 1 and 2, respectively. Further, we have analyzed the legacy mechanisms and the current state of the art towards their utility for 5G and B5G networks in Tables 3 and 4, respectively. However, we have observed that gaps in fulfilling the requirements still persist. Concretely, we have demonstrated that none of the strategies evaluated satisfy the reliability, flexibility and scalability criteria in their entirety. 
Hence, to be able to design and develop a holistic MM mechanism, it is of substance to our study to understand the challenges/questions that persist. We consolidate, from earlier works in literature and the discussion in Sections 2-5, these key challenges/questions in the text that follows.}

\subsection{\textcolor{red}{Challenges}}
\subsubsection{\textcolor{red}{Handover Signaling}}

\textcolor{red}{Even after the release of 3GPP specifications for 5G \cite{Specification2017a}, HO signaling is still a challenge. Hence, reducing HO signaling to ensure system scalability and reliability will be one of the key challenges. Certain studies such as \cite{Jain2019} have provided methods to help overcome this challenge, and hence, can be actively pursued by the research and industrial community.}

\subsubsection{\textcolor{red}{Network Slicing}}

\textcolor{red}{Network slices have been defined to ensure different service types are served according to their own resource demands. Hence, it will be a key challenge to design MM strategies that either jointly take into account the requirements of multiple network slices or provide individual solutions for each network slice.} 

\subsubsection{\textcolor{red}{Integration framework for MM solutions}}

\textcolor{red}{The state of the art and 3GPP specifications ensure to some extent the provision of flexibility, reliability and scalability for 5G MM solutions, as discussed earlier. However, since these solutions function at different sections of the network (Figure 2), the challenge will be to design them such that collectively they ensure the appropriate levels of flexibility, scalability and reliability in MM mechanisms to cope with the diversity in mobility profiles and applications the devices will access. Also, a part of this challenge will be to ensure that the CAPEX and Operating Expenditure (OPEX), owing to the architectural (software or hardware) transformations stemming from these redesigned MM mechanisms, are manageable.}  

\subsubsection{\textcolor{red}{Ensuring Context Awareness}}

\textcolor{red}{Context based MM solutions accounting for factors such as network load, user preference, network policy, mobility profiles, etc., to ensure best possible provision of requested QoS will be important. The criticality of this challenge is enhanced by the fact that, low computational complexity whilst executing these solutions will be of the essence to meet the strict latency constraint requirements.}


\subsubsection{\textcolor{red}{Architectural Evolution Costs}}    

\textcolor{red}{SDN and edge cloud capabilities will be important for enhancing the user experience during mobility, as discussed in Section 5. However, a key challenge will be to ensure appropriate scalability while maintaining a manageable CAPEX and OPEX.}

\subsubsection{\textcolor{red}{Frequent Handovers}}
\textcolor{red}{Reducing frequent handovers, ping-pong effects and devising an optimized HO strategy will still be a key challenge, given the dense and heterogeneous future network environment. This is further exacerbated by the fact that current methods, such as IEEE 802.21 and 3GPP specifications, fail to integrate cellular and non-3GPP networks effectively for seamless HO between them. For example, while methods such as LWA have been explored extensively \cite{Ratasuk2014, Alkhansa2014}, an effective handover methodology between 3GPP and non-3GPP networks still remains elusive.}

\subsubsection{\textcolor{red}{Security}}
\textcolor{red}{An important challenge for ensuring service continuity and seamless mobility in an extremely dense and heterogeneous network environment, such as 5G and beyond networks, will be to ensure that security related tasks, such as authenticating the user as well as the network, be completed as efficiently as possible. By efficiently here we mean that the authentication should guarantee a required level of security whilst provisioning low computational complexity \cite{Ferrag2017} as well as latency \cite{JawadAlam2018}. Again this task will become even more critical in scenarios where mobility occurs between 3GPP to non-3GPP networks.}

\subsubsection{\textcolor{red}{Energy Efficiency}}

\textcolor{red}{Given that one of the goals of 5G is to ensure enhanced battery lives for the devices, it will be a critical component for 5G MM services to ensure that the mobility of the devices is handled in an energy efficient way \cite{Qiao2017}. Additionally, 5G MM services will also need to ensure that the energy footprint goal for 5G networks is achieved via techniques such as smart AP selection methodologies \cite{Habbal2017} and reduced CN signaling \cite{Jain2019}. By smart AP selection methodologies we refer to being able to not only account for the user energy consumption over the course of its mobility, but also accounting for the energy consumed whilst performing such selections.}

\subsubsection{\textcolor{red}{Meta-surface Reconfiguration for mobility support}}

\textcolor{red}{For the B5G networks, finding the optimal configuration of meta-surfaces during mobility related scenarios will be challenging. This is because, the physical characteristics of the surfaces will have to be altered rapidly so as to have the signals arriving at the user in a constructive manner.}

\subsubsection{\textcolor{red}{Beyond 5G Network: Handovers}}

\textcolor{red}{A fundamental question that will be posed in B5G is -- how frequently and when will the handovers be needed? The reason this question is a challenge because, up until now the rate of power loss in an urban environment is characterized by a $R^4$ factor (where $R$ is distance between the transmitter and receiver) given the destructive interference encountered. However, with programmable environments, according to \cite{Renzo2019}, this decay will now be similar to the free space scenario, i.e., $R^2$, since all signals can be modulated in phase and polarization to interfere at the receiver in a constructive manner. And so, in mobile environments, the power decay will not be significant even at distances further away. Hence, the handover triggering methods and their execution procedure need to be revisited as currently they do not expect such a reliable behavior from the channel.}

\subsubsection{\textcolor{red}{Beyond 5G Network: Protocol stack}}

\textcolor{red}{A next fundamental question posed in B5G, with reference to meta-surfaces, is: What is the impact on the existing layers? The reason this question is a challenge because, the MAC, Radio Link Control (RLC), PDCP and TCP layers, they all have error control, packet re-ordering, transmission repeat request and other reliability control mechanisms in-built. These were designed keeping in mind that the environment is unreliable and randomly varying. However, with programmable surfaces the environment will be much more deterministic and reliable. Thus, there arises a case for either eliminating/modifying some of these layers (for example, a lightweight version of TCP may be utilized , as the channel is deterministic and the probability of having lost packets due to error or timeout is significantly lower since the multipaths can be redirected to interfere constructively at the receiver by the meta-surfaces, or the User Datagram Protocol (UDP) can be utilized with much more reliability), which play a critical part in MM procedures, or revisiting their original implementation to adapt to these programmable environments.}

\subsubsection{\textcolor{red}{Dynamic Network Topology}}
\textcolor{red}{In terms of user association for B5G networks, the challenge will now not be to just choose an AP with the best SINR/RSSI/RSRP/RSRQ, but it will rather be to choose or program an AP/programmable surface configuration/drone, depending on the user mobility, location and coverage from these sources. While it still reduces to the problem presented for 5G networks, the increased dimensionality and heterogeneity of the problem will provide formidable challenges to existing methods.}

\subsubsection{\textcolor{red}{Edge Node configuration in B5G networks}}

\textcolor{red}{Edge nodes' placement for supporting user mobility will also be challenged. This is so because the possibility of supporting better QoS over longer distances can reduce the requirements for service replication/service migration. This is a consequence of the fact that the handovers would be impacted given the programmability of the environment and the squared decay instead of a fourth power decay in the received signal power.}

\subsubsection{\textcolor{red}{IP address continuity}}

\textcolor{red}{The vision for near zero latency by 3GPP \cite{3GPP261} necessitates that E2E link continuity is ensured given any network and mobility scenario. Hence, maintaining IP address continuity during mobility events will remain a critical challenge as the complexity of the networks increases in 5G and B5G.\\ }
	
\noindent \textcolor{red}{The aforementioned key challenges define the technology gap towards fulfilling the MM governing parameters listed in Table 2. In the following subsection we list the potential solutions that can fill this technology gap.}


\subsection{Potential Solutions}
\subsubsection{\textcolor{red}{Smart CN signaling}}
\textcolor{red}{Utilizing the properties of SDN, the signaling performed within the CN for handover and re-routing purposes can be optimized further. This will enable more scalability and better support to users with high mobility. Concretely, techniques such as graph theory, Machine Learning \cite{Prieto2017} as well as the recently established intelligent Information Elements (IE) mapping methods \cite{Jain2019}, etc., can enable faster and efficient CN signaling, as mentioned above. Here by efficiency we imply that the transmission cost, processing cost and other CN signaling related metrics \cite{Jain2019} are reduced/optimized.}

\subsubsection{\textcolor{red}{On demand MM}}
\textcolor{red}{Given the functional requirements (Section 2), legacy methods (Section 4) and the state of the art (Section 5), on demand MM strategies (such as \cite{Jain2017}) will allow future MM mechanisms to serve users with different mobility profiles, accessing different services and accessing networks with differing loads, more effectively. As an example, slice based MM strategies can enable independent strategies for the various network slices that the 5G networks will serve. This will help cater to the different network slices according to their mobility demands, and avoid the sub-optimal \emph{one size fits all} approach.}

\subsubsection{\textcolor{red}{Deep learning}}
\textcolor{red}{Learning network parameters such as network load, congestion statistics at access and core network, user mobility trends, etc., enable the network to devise effective and optimal MM strategies for a highly dynamic network environment such as that in 5G and B5G networks. Hence, deep learning methods such as reinforcement learning can assist in such tasks.}

\subsubsection{\textcolor{red}{SDN-NFV integrated DMM}}
\textcolor{red}{DMM facilitates the distribution of MM functionality throughout the network and avoiding single MM anchors, which consequently assists in alleviating issues such as SPoF and congestion. Note that, SDN and NFV will assist in DMM as network programmability facilitates fast switching while the user/device transits through the network.}

\subsubsection{\textcolor{red}{D2D CP-DP extension}}
\textcolor{red}{D2D clustering and support for communication with devices in such clusters has been formalized since 3GPP Release-13. Thus, through an extension of CP-DP capabilities of the current D2D framework, i.e., by utilizing the relaying strategies for CP/DP information, handover performance for devices migrating within the network and in such clusters can be enhanced. Further, policy based methods, which take into account the presence of D2D communications between vehicles and other V2X scenarios, will also enable future MM mechanisms to serve the complex scenarios that will prevail in 5G and B5G networks better.}

\begin{table*}
    \color{red}
    \caption{Mapping potential solutions to MM challenges} 
    \centering
    \begin{tabular}{|>{\centering\arraybackslash}m{2cm}|>{\centering\arraybackslash}m{2.5cm}|m{11cm}|>{\centering\arraybackslash}m{1.5cm}|} \hline
        \textbf{Challenges} & \textbf{Recommended Potential Solutions} & \centering\arraybackslash\textbf{Comments} & \textbf{Param. Satisfied$^*$} \\ \hline
         Handover Signaling & Smart CN Sig. \& SDN-NFV integ. DMM & In addition to the existing strategies, a smart CN signaling method, such as that in \cite{Jain2019} will assist in relieving the handover signaling load significantly. DMM strategies will assist in decentralization of MM anchors and hence, more reliability in mobile environments & RL3, RL5, SL1 -- SL4 \\ \hline
        Network Slicing & On demand MM & An on demand strategy will assist the network slices to assist in provisioning tailor made mobility solutions for the corresponding tenants & FL1, FL5\\ \hline
        Integration framework for MM solutions & \textit{Design} & This is a design challenge and hence, should collectively take into account all the other non-design challenges as well as other necessary factors, such as efficacy and delays & SL5\\ \hline
        Ensuring Context Awareness & On demand MM & It will ensure that the user, network and application context is taken into account and appropriate MM solution is provisioned as and when needed & FL5\\ \hline
        Architectural Evolution Costs & \textit{Design} & This is a design challenge and hence, should collectively take into account all the other non-design challenges as well as other necessary factors, such as cost of infrastructure & SL5\\ \hline
        Frequent Handovers & Deep learning & Learning the network conditions, mobility profiles and the corresponding impact on the handovers is a complex task. Deep learning can help predict/estimate valuable system parameters, such as SINR, to avoid the frequent handover condition via appropriate AP-user association & RL1, RL2, FL2, FL3, FL4 \\ \hline
        Security & Smart CN Signaling & Effective CN signaling will assist in maintaining/migrating security context when required, thus reducing the latency as well as complexity to ensure the same & RL2, SL3\\ \hline
        Energy Efficiency & Deep learning and Smart CN Signaling & Whilst deep learning methodologies can in general provision an optimal solution for handling user mobility whilst adhering to the energy constraints, smart CN signaling, via reduction in signaling messages during mobility, can enhance energy efficiency of the MM strategy  & SL1 \\ \hline
        Meta-surface Reconfiguration for mobility support & Deep learning & Based on the user mobility deep learning algorithms can assist in understanding how the meta-surface configurations have to be adjusted so as to ensure the requested QoS for the users & RL1,RL2 and FL3  \\ \hline
        B5G: Handovers & Smart CN Sig., Serv. Cont. through Edge Comp. \& D2D CP-DP Ext. & Edge compute platforms can assist in faster and effective handover decisions, given their capability to provision compute power closer to the access network. Smart CN signaling can assist in efficient and low latency handover signaling in the CN. D2D networks can assist in extended coverage and hence, smoother handovers  & RL2, RL4, FL3, SL1 -- SL4\\ \hline
        B5G: Protocol Stack & \textit{Design} & This is a design challenge and hence, should collectively take into account all the other non-design challenges as well as other necessary factors, such as efficacy and delays& SL5\\ \hline
       Dynamic Network Topology & Deep learning & The ability to understand complex associations will make deep learning methodologies essential in determining the optimal user-AP association in an increasingly dynamic and multi-dimensional network, such as the B5G networks & RL1, RL2, FL3, FL4\\ \hline
        Edge Node Configuration in B5G Networks & \textit{Design} & This is a design challenge and hence, should collectively take into account all the other non-design challenges as well as other necessary factors, such as efficacy and infrastructure cost& SL5\\ \hline
        IP address continuity & Clean Slate Methods & Given their ability to resolve destinations based on names and not the IP address, clean slate methods can assist in maintaining a single IP address throughout with respect to the destination server & RL2, RL4 \\ \hline
        \multicolumn{4}{l}{$^*$ Details regarding the parameters and the requirements that they help satisfy are provided in Table 2.} 
    \end{tabular}
    
\end{table*}

\subsubsection{\textcolor{red}{Service Continuity through Edge Computing}}
\textcolor{red}{For serving fast moving users, such as vehicles, and satisfying their latency and bandwidth requirements, edge computing solutions for MM will play a major role in 5G and B5G networks \cite{Boban2017}. And while service migration strategies will play a critical role in ensuring seamless connectivity, a fine balance between service replication and service migration will help mitigate the multitude of challenges that arise for such strategies. Further, given that users might crossover to other PLMNs during the duration of mobility \cite{Report2018}, which can lead to a change in the edge cloud that serves them, effective service migration strategies will greatly enhance the QoS during mobility.}
\begin{figure*}
	\centering
		\includegraphics[scale=0.38]{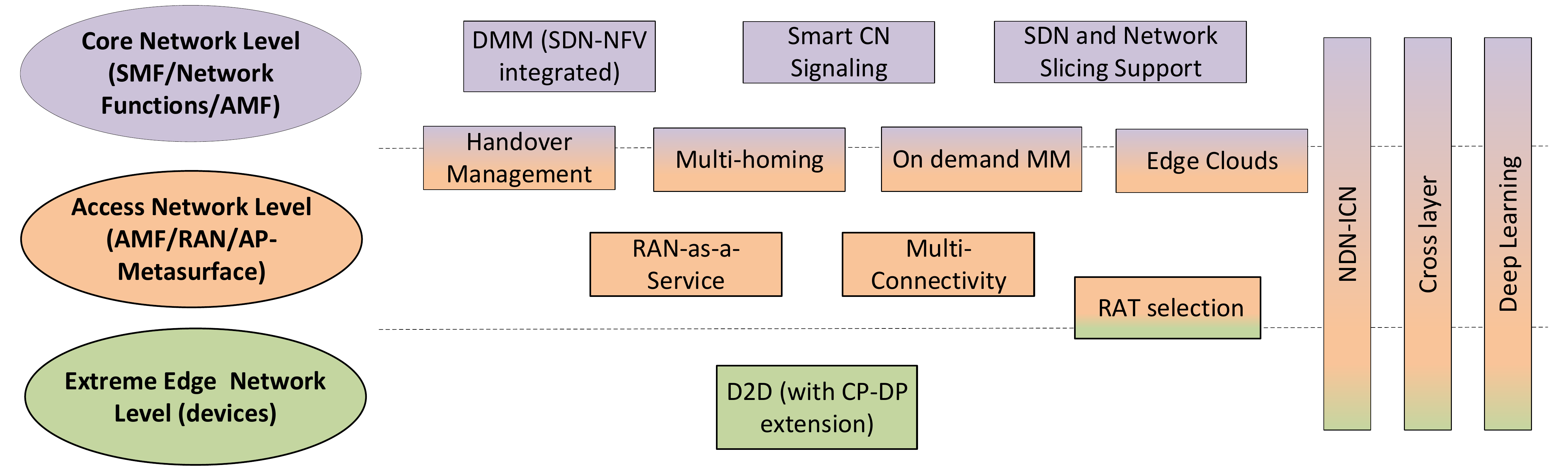}
	\caption{Proposed 5G and beyond MM framework.}
\end{figure*}

\subsubsection{\textcolor{red}{Clean Slate Methods}}
\textcolor{red}{Current networks rely on resolving the IP addresses of the hosts for the applications requested by the users. However, such a resolution can lead to delays \cite{Zhang2014}. And so, Information Centric Networking (ICN), and specifically Named Data Networking (NDN) paradigm, avoid this process thus making the network more flexible and faster. Additionally, with the proposition of having in-network caching, ICN and NDN paradigms enable caching capabilities near the users.} 
    
\textcolor{red}{Another class of such clean slate methods is MobilityFirst \cite{Raychaudhuri2012}. In MobilityFirst, a new paradigm to networking, like in ICN and NDN, has been proposed. In this paradigm, IP based resolution of nodes has been deprecated and name based resolution is proposed. Further, concepts similar to ICN and NDN, such as in-network caching etc., have also been proposed. Additionally, and different to the ICN-NDN paradigm, ensuring security in a fully dynamic scenario has been considered as one of the guiding principles of MobilityFirst. Further, MobilityFirst also introduces support for migration of entire networks and not just the end nodes.}
    
\textcolor{red}{Consequently, such methods together can provision more scalable, flexible and reliable MM strategies.\\}

\noindent \textcolor{red}{And so, up until now in this section, we have highlighted the multiple challenges that the 5G and beyond MM mechanism will face, given our qualitative evaluation for legacy and current state-of-the art methods in Sections 2-5. We have then provisioned a brief discussion on the potential solutions that can assist in addressing these challenges. We illustrate a novel mapping between these challenges and potential solutions in Table 5. Additionally, we have also listed the parameters for the qualitative analysis (and hence the requirements specified in Table 1) that they satisfy. This, as a result, reinforces the completeness of our current study. Hence, in the next subsection, utilizing the inferences from Sections 2-5 and Table 5, we propose a framework for 5G and beyond MM.}

\subsection{\textcolor{red}{Proposed 5G and beyond MM framework}}

\textcolor{red}{We utilize the earlier established classification process for the current state-of-the-art strategies to define our vision for 5G and beyond MM in Figure 3. Concretely, we have categorized the MM mechanisms as \textit{Core Network level}, \textit{Access Network level} and \textit{Extreme Edge Network level}, depending on where they will be creating an impact on/from. The specific entities (based on the 5G architecture illustrated in Figure 2), to which these aforesaid levels correspond to, have also been mentioned in Figure 3.}   

\textcolor{red}{To elaborate, the core network strategies encompass the DMM, SDN and Network slicing paradigms to provision the necessary reliability, flexibility and scalability from a more global perspective. Additionally, the aforesaid core network strategies need to be well complemented with an efficient CN signaling strategy. Next, handover management, on-demand MM, IPv6 multi-homing and Edge cloud related MM strategies will be enacted  not only in the core network or the access network level, but jointly at both levels thus provisioning the necessary flexibility and reliability. Further, RAN-as-a-Service and Multi-connectivity provisions at the access network level will assist in utilizing the multiple RATs and APs effectively. Moreover, it is envisioned that the RAT selection process maybe either at the access network or at the device level. The D2D techniques, on the other hand, are expected to provide added assistance for mobility at the device level through DP as well CP functionality.}

\textcolor{red}{Complementing these mechanisms, NDN-ICN support will be provisioned at all levels, thus assisting in maintaining IP addresses/prefixes during mobility whilst resolving destinations via names. Note that, traditional IP address/prefix allocation strategies are not intended to be changed. Instead, the NDN-ICN concept provisions an over-the-top assistance. Further, the cross layer strategies, as the name suggests, will spawn across the multiple levels and enact policies, utilizing the available information at each of these levels, which assist in optimal MM related decisions across the network. Lastly, the deep learning strategies will again assist across the multiple levels by learning the complex features about the network context, user mobility and overall QoS requirements, and formulating effective MM related decisions.}

\textcolor{red}{Hence, given that we utilize the potential solutions for overcoming the technology gap, specified in Section 6.2, alongside certain strategies from the state of the art and legacy MM mechanisms, specified in Sections 4 and 5, it can be inferred from Tables 2-5 that our proposed framework will satisfy all the parameters for the reliability, flexibility and scalability criteria. Consequently, it can be stated that the proposed framework in Figure 3 will also satisfy all the requirements as defined in Table 1, thus provisioning a holistic solution. With this vision, in the following section we summarize the main findings of this article and conclude this paper.}

\section{Conclusions}

Given the complexity of future network scenarios, i.e., 5G and B5G, a full view of the MM strategies, their capabilities, the persistent challenges and the possible solutions to them, will enable the research community to design better MM strategies. 

\textcolor{red}{In this paper, through Section 2 and Table 1, we firstly presented the important functional requirements and design criteria to be considered when devising 5G and B5G MM solution. We then presented the multiple parameters that the future MM mechanisms needs to satisfy for each of the evalutation criteria, i.e., scalability, flexibility and reliability, in Section 3 and Table 2. Next, from our discussions in Section 4 it is clear that the legacy MM solutions fail in provisioning scalability, flexibility and reliability simultaneously. Nevertheless, the current standards and research efforts explored in Section 5 are promising as they provide enhanced capabilities towards future MM solutions. We have summarized these conclusions effectively in Tables 3 and 4. And as a consequence, through this qualitative analysis the various benefits and shortcomings of the legacy and the current state of the art mechanisms, studied in this paper, can be understood easily by the research community. Subsequently, we established that none of the mechanisms fulfill the complete 5G and beyond MM mechanism requirements. }

\textcolor{red}{And so, it is evident that a holistic MM mechanism for 5G and B5G networks remains elusive. Thus, certain challenges that will still persist for the design, development and deployment of future MM mechanisms have been detailed in this paper in Section 6.1. Furthermore, we have provided a concise discussion on the potential MM strategies that the research community can explore so as to solve these persistent challenges and the technological gaps they present, in Section 6.2. Following this, we have also provisioned a novel mapping between the potential strategies and the persistent challenges in Table 5, thus highlighting the efficacy of our current study. Based on the inferences drawn, we have concluded our study by provisioning a novel framework for the 5G and beyond MM strategies through Section 6.3 and Figure 3.} 

\section*{References}

\bibliographystyle{ieeetran}

\end{document}